\documentclass[a4paper,11pt]{article}
\usepackage{setspace}
\usepackage[left = 2cm, right=2cm, top=3cm, bottom=3cm]{geometry}
\usepackage[pdftex]{graphicx}
\usepackage{cite}
\usepackage{bm}
\usepackage{float}
\usepackage{amsmath,amssymb,latexsym}
\usepackage{color}
\usepackage[T1]{fontenc}
\usepackage{wasysym}
\usepackage{siunitx}
\usepackage{setspace}
\usepackage{relsize}
\usepackage{tikz}
\usepackage{authblk}
\usepackage[symbol]{footmisc}
\usepackage{lineno}
\usepackage{stfloats}

\usepackage{cite}
\usepackage{hyperref}
\begin{document}
	
	
	\renewcommand{\figurename}{Fig.}
	
	\title{\color{blue}\textbf{Emergence of transient reverse fingers during radial displacement of a shear-thickening fluid}}
	\author[1, $\dagger$, $\P$ ]{Palak}
	\affil[1]{\textit{Soft Condensed Matter Group, Raman Research Institute, C. V. Raman Avenue, Sadashivanagar, Bangalore 560 080, INDIA}}
	\author[1, $\ddagger$, $\P$]{Vaibhav Raj Singh Parmar}
        \author[1, $\S$]{Sayantan Chanda}
	\author[1,*]{Ranjini Bandyopadhyay}
	\date{\today}
        \footnotetext[5] {Palak and Vaibhav Raj Singh Parmar contributed equally to this work.}
	\footnotetext[2]{palak@rri.res.in}
	\footnotetext[3]{vaibhav@rri.res.in}
        \footnotetext[4]{sayantanc@rrimail.rri.res.in}
	\footnotetext[1]{Corresponding Author: Ranjini Bandyopadhyay; Email: ranjini@rri.res.in}

	\maketitle
	\begin{abstract}

A highly sheared dense aqueous suspension of granular cornstarch particles displays rich nonlinear rheology. We had previously demonstrated the growth and onset of interfacial instabilities when shear-thinning cornstarch suspensions were displaced by a Newtonian fluid, and had suggested methods to maximise displacement efficiency~[Palak, R. Sathayanath, S. K. Kalpathy and R. Bandyopadhyay, Colloids Surf. A Physicochem. Eng. Asp., 629 (2021) 127405]. In the present work, we explore the miscible displacement of a dense aqueous cornstarch suspension  {in its discontinuous shear-thickening regime} in a quasi-two-dimensional radial Hele-Shaw cell. We systematically study the growth kinetics of the inner interface between water and the cornstarch suspension, and also of the outer interface between the suspension and air. In addition to the growth of interfacial instabilities at the inner interface, we observe a transient withdrawal of the suspension and the formation of fingering instabilities at the outer interface. We demonstrate that these `reverse fingering' instabilities are extremely sensitive to the injection flow rate of water, the gap of the Hele-Shaw cell and the concentration of the displaced cornstarch suspension,  {and emerge irrespective of immiscibility between the fluid pair. We believe that as the cornstarch suspension dilates due to the high shear rate imposed by the displacing fluid, the outer suspension-air interface responds with a restoring force, resulting in the penetration of air into the suspension and the formation of reverse fingers. We note that the growth of reverse fingers significantly reduces the displacement efficiency of the suspension. Finally, we demonstrate a correlation in the growth of inner and outer interfacial patterns by computing the velocity with which stresses propagate in the confined dense suspension. Our findings are useful in understanding the flow of granular materials through constrained geometries and can be extended to study stress propagation in shear-thickening materials due to a sudden imposition of high shear rate, such as in impact behaviour.}
	\end{abstract}
	
	\definecolor{black}{rgb}{0.0, 0.0, 0.0}
	\definecolor{red(ryb)}{rgb}{1.0, 0.15, 0.07}
	\definecolor{darkred}{rgb}{0.55, 0.0, 0.0}
	\definecolor{blue(ryb)}{rgb}{0.01, 0.2, 1.0}
	\definecolor{darkcyan}{rgb}{0.0, 0.55, 0.55}
		\definecolor{navyblue}{rgb}{0.0, 0.0, 0.5}
		\definecolor{olivedrab(web)(olivedrab3)}{rgb}{0.42, 0.56, 0.14}
			\definecolor{darkraspberry}{rgb}{0.53, 0.15, 0.34}
				\definecolor{magenta}{rgb}{1.0, 0.0, 1.0}
		
	\newcommand{\blsquare}{\textcolor{black}{\small$\blacksquare$}}
	\newcommand{\hlsquare}{\textcolor{darkred}{\small$\square$}}
		\newcommand{\redtraingle}{\textcolor{magenta}{\small$\triangle$}}
			\newcommand{\oolive}{\textcolor{olivedrab(web)(olivedrab3)}{\large$\circ$}}
			\newcommand{\purpletraingle}{\textcolor{darkraspberry}{\small$\triangledown$}}
		\definecolor{lime(web)(x11green)}{rgb}{0.0, 1.0, 0.0}		
	\newcommand{\bltriangle}{\textcolor{black}{\small$\triangleup$}}
	\newcommand{\rcircle}{\textcolor{red(ryb)}{\large$\bullet$}}
	\newcommand{\rtraingle}{\textcolor{red(ryb)}{\small$\triangledown$}}
	\newcommand{\wine}{\textcolor{darkred}{\large$\bullet$}}
	\newcommand{\cyan}{\textcolor{darkcyan}{\large$\bullet$}}
	\newcommand{\owine}{\textcolor{darkred}{\large$\circ$}}
	\newcommand{\ogreen}{\textcolor{lime(web)(x11green)}{\large$\circ$}}
	\newcommand{\redcircle}{\textcolor{red}{\large$\circ$}}
	\newcommand{\ocyan}{\textcolor{darkcyan}{\large$\circ$}}
	\newcommand{\blue}{\textcolor{blue(ryb)}{\large$\bullet$}}
		\newcommand{\blbullet}{\textcolor{navyblue}{\large$\bullet$}}
		\newcommand{\olbullet}{\textcolor{olivedrab(web)(olivedrab3)}{\large$\bullet$}}
	\newcommand{\hollowblue}{\textcolor{blue(ryb)}{\large$\circ$}}
	\newcommand{\black}{\textcolor{black}{\large$\bullet$}}
	\newcommand{\hollowblack}{\textcolor{black}{\large$\circ$}}
		\newcommand{\bltria}{\textcolor{black}{\small$\triangle$}}
			\newcommand{\rtrai}{\textcolor{red(ryb)}{\large$\triangledown$}}
			\newcommand{\rline}{\raisebox{2pt}{\tikz{\draw[-,red!40!red(ryb),solid,line width = 2.5pt](0,0) -- (5mm,0);}}}

	\section{\label{Intro}Introduction}
	Shear-thickening (ST), which refers to the enhancement of the apparent viscosity of a suspension as an externally imposed shear stress or shear rate is progressively increased~\cite{shearthickening,shearthickening2}, is observed in  materials such as aqueous suspensions~\cite{Silica,doi:10.1073/pnas.2203795119} and granular mixtures~\cite{Fall2010shear,Granular2}. Dense granular suspensions of polydisperse irregular-shaped cornstarch particles in water were reported to exhibit shear-thinning and shear-thickening properties depending on the applied shear stress~\cite{peters2016direct} and particle volume fraction~\cite{wagner2009shear}. An aqueous cornstarch suspension subjected to low shear rates exhibits shear-thinning behaviour due to several factors, such as the organisation of suspended particles along the flow, a constant hydrodynamic viscosity contribution due to viscous stresses~\cite{wagner2009shear} and an entropic contribution from random particle collisions~\cite{shearthinning,CSconcentration}. As the externally imposed shear rate is increased, a dense cornstarch suspension shows a continuous shear-thickening (CST) regime which is followed by discontinuous shear-thickening (DST) behaviour~\cite{Fall2010shear}. At even higher imposed shear stresses, dense cornstarch suspensions display a shear jamming (SJ) state in which the suspension no longer flows but instead behaves like a solid~\cite{peters2016direct}. The increase in the bulk viscosity of the suspension in the ST regime is attributed to hydrodynamic~\cite{hydrodynamics,wagner2009shear} and frictional~\cite{frictional,WyartPhys} interactions between the constituent particles. The CST regime is characterised by well-defined highly stressed dynamic regions which propagate in the shearing direction and span an increasingly larger fraction of the suspension as the applied stress is increased~\cite{CST}. Recent simulations of dense suspensions have shown that inter-particle connectivity induced by the formation of force networks determines the rheological response of the suspension~\cite{Jamali}. It is reported that the DST regime has more constrained force networks connected via multiple particle contacts when compared to the CST regime which is characterised by less constrained force networks having single particle-particle contacts. The number density of these frictional contacts increases with increasing shear rates until the SJ state is reached~\cite{peters2016direct}. Measurements of the first normal stress difference $N_1$~\cite{macosko1994rheology} can indirectly shed light on the underlying inter-particle interactions contributing to the generation of large stresses in concentrated suspensions~\cite{normal,normal1}. While negative values of $N_1$ suggest hydrodynamic or lubrication effects in the suspension, positive $N_1$ values indicate the presence of inter-particle friction~\cite{normal}.

Material transport in industrial processing may cause interfacial instabilities, which can affect production efficiency. One such interfacial instability is the Saffman-Taylor instability, which involves the development of an intricate interface when a less viscous fluid displaces a more viscous one~\cite{saffman1958penetration,Homsy,pinilla2021}. The viscous fingering (VF) instability was initially identified in oil recovery fields when water was injected under high pressures into a porous medium~\cite{orr1984use,oilrecovery}. Since then, many studies have been performed to understand this phenomenon using a Hele-Shaw (HS) geometry~\cite{pinilla2021} which comprises two glass plates separated by a narrow gap.   {In general, the fluid flow in a HS geometry is mathematically analogous to flow in a porous medium~\cite{saffman1958penetration}.} Several factors affect the growth of these instabilities, for example, the wettability of the fluid pair~\cite{ESLAMI201925}, the gap of the HS cell~\cite{Van_gap} and fluid rheology~\cite{Linder2000Viscous,Associating1993zhao,buka1986transitions,ozturk2020flow}. Interfacial instabilities have been systematically investigated in non-Newtonian fluids with exotic nonlinear rheological responses such as shear-thinning, shear-thickening and non-zero yield stresses. VF in non-Newtonian fluids, for example, in liquid crystals~\cite{buka1986transitions,Zhang2021Structures}, polymers~\cite{Linder2000Viscous,Associating1993zhao}, colloidal suspensions~\cite{criterion2020divoux,Lemaire1991,PhysRevFluids.3.110502,Kawaguchi,Kawaguchi2,PALAK2022100047}, emulsions~\cite{Viscous2004kawaguchi} and granular materials~\cite{ozturk2020flow,cheng2008towards,PALAK2021127405,sandnes2011patterns}, have been studied both experimentally and numerically during the past few decades.  {Interfacial instabilities in unfavourable viscosity conditions, driven by the presence of particles at the propagating fluid front, have also been studied in the literature~\cite{PhysRevLett.117.034501,PhysRevLett.118.074501}.} Experiments involving the displacement of shear-thickening propylene glycol (PPG)-silica suspensions by air~\cite{Kawaguchi} demonstrated that the finger velocity deviates from the prediction of the modified Darcy's law~\cite{modDarcy} as the injection pressure is increased. When a silica suspension was displaced by air at a shear rate exceeding the critical value necessary to initiate suspension shear thickening, a transition from a stable pattern to VF instability was observed~\cite{Kawaguchi2}. Another report on the displacement of shear-thickening cornstarch suspensions by air reported an excellent correlation between suspension rheology and the observed interfacial pattern morphologies~\cite{ozturk2020flow}. This work demonstrated that interfacial pattern morphologies change with increasing injection pressures as the cornstarch suspension transitions from one flow regime to another. While earlier research work focussed on the miscible displacements of shear-thinning cornstarch suspensions~\cite{PALAK2021127405} and the immiscible displacements of shear-thickening suspensions~\cite{Kawaguchi,Kawaguchi2,sandnes2011patterns,ozturk2020flow}, the miscible displacement of shear-thickening suspensions has never been investigated experimentally to the best of our knowledge.

In our previous report on the miscible displacement of cornstarch suspensions in the shear-thinning regime, we showed that increasing elasticity of the suspension and viscosity ratio of the fluid pair resulted in suppression of interfacial instability~\cite{PALAK2021127405}. In the present work, we explore the influence of the shear-thickening rheology of a dense cornstarch suspension (displaced outer fluid) on the onset and growth of interfacial instabilities during its radial displacement by water (inner displacing fluid) in a quasi-two-dimensional Hele-Shaw (HS) cell. While the existing literature on the study of instabilities focusses exclusively on the propagation of the inner fluid front at the interface between the inner displacing and outer displaced fluids, our present work observes two growing interfaces simultaneously in a single experiment, $viz.$, the inner interface between water and displaced cornstarch suspension, and the outer interface between cornstarch suspension and air (outermost fluid). We observe a transient withdrawal of the cornstarch suspension and the evolution of reverse fingers at the outer interface during displacement of the suspension by water  {for injection flow rates imposed above a critical value}. We demonstrate that  { the generation of reverse fingers is insensitive to the miscibility of the displacing fluid} but depends sensitively on injection flow rate, concentration of the cornstarch suspension and gap of the HS cell. We quantify this novel reverse fingering phenomenon by estimating the perimeter of the pattern ($\Delta P_{out}$), the number of reverse fingers ($N_{rf}$) and the average spacing between these fingers ($\lambda$) at the outer interface.  {Our normal stress measurements strongly suggest that the displaced cornstarch suspension lies in the discontinuous shear-thickening (DST) regime. As a result of the high shear rate imposed by the displacing fluid, the cornstarch suspension is expected to dilate, such that the CS particles that are constrained to assemble in force networks penetrate the outer suspension-air interface~\cite{brown2012role, maharjan}. Such surface deformation results in a restoring force, with air propagating into the suspension in a radially inward direction, giving rise to the observed reverse fingers. Finally, we report a clear correlation in the interfacial dynamics at the inner and outer interfaces and show that emergence of reverse fingers at the outer interface reduces efficiency of displacement of the cornstarch suspension.}

\section{\label{em}Materials and Methods}
\begin{figure}[h!]
	\includegraphics[width= 3.0in]{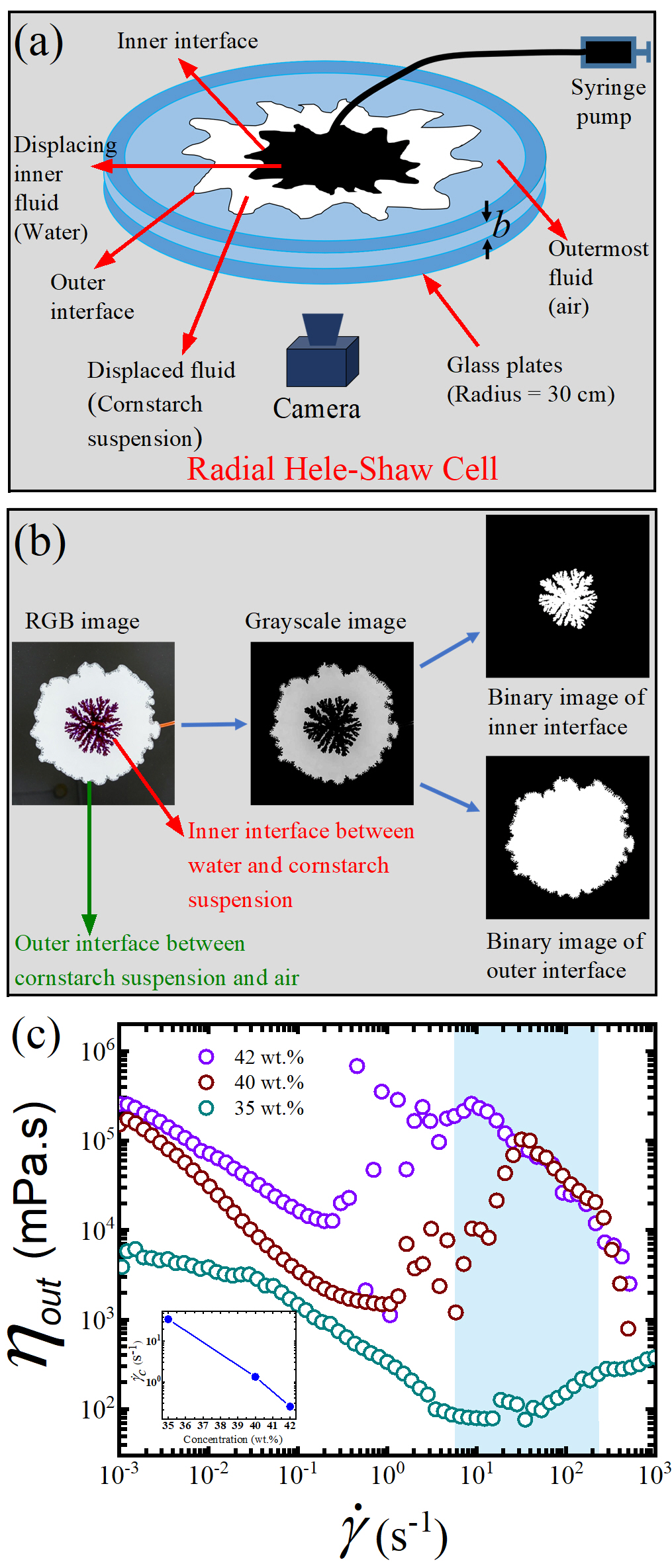}
	\centering
	\caption{\label{fig:hssem}{\bf The experimental setup, binarization procedure, and rheological measurements.} \textbf{(a)} Schematic illustration of a radial Hele-Shaw (HS) cell. \textbf{(b)} Binarization steps for detecting the inner interface between water (inner displacing fluid) and cornstarch (CS) suspension (outer displaced fluid), and the outer interface between CS suspension and air (outermost fluid). \textbf{(c)}  {Shear rate dependent viscosities of dense cornstarch suspensions, $\eta_{out}$, vs. shear rates, $\dot{\gamma}$, for different cornstarch particle concentrations at a rheometer plate separation of 300 $\mathrm{\mu}$m.} The region highlighted in blue indicates the shear-thickening flow regime of a 40 wt.\% cornstarch suspension. Inset shows the variation in critical shear rate $\dot{\gamma}_c$ for the onset of shear-thickening as a function of concentration of the CS suspension.}
\end{figure}
	
A radial Hele-Shaw (HS) cell setup~(Fig.~\ref{fig:hssem}(a)), consisting of two circular glass plates, each of radius 30~cm and thickness 10~mm, is used to study the displacement of a dense cornstarch suspension by water. Teflon spacers of thicknesses 170 $\mathrm{\mu}$m, 300 $\mathrm{\mu}$m, 500 $\mathrm{\mu}$m and 800 $\mathrm{\mu}$m are used to maintain a constant gap between the glass plates. The fluids are injected with a syringe pump (NE-8000, New Era Pump Systems, USA) through a 3 mm hole drilled at the centre of the top plate. In our experiments, density matched cornstarch suspensions ($\rho$ = 1.59 $\mathrm{g/cm^3}$) are prepared by homogeneously mixing cornstarch powder (Sigma-Aldrich) in a 55 wt.\% aqueous solution of cesium chloride CsCl (ReagentPlus\textsuperscript{\textregistered}, Sigma-Aldrich)~\cite{Browndynamic,merktpersistent} using a magnetic stirrer (1 MLH, Remi Equipments Ltd., Mumbai), followed by ultrasonication (USC 400, ANM Industries Pvt. Ltd.). The sample is left undisturbed for 24 hours to ensure uniform hydration of the cornstarch particles~\cite{CSconcentration}.

To perform displacement experiments, the homogeneous cornstarch suspension is first loaded in the radial HS cell until its boundary reaches approximately 10 cm from the injection point.  {Different filling lengths of CS suspension, $viz.$, 5 cm, 7.5 cm and 15 cm, are also used in additional experiments.} After loading the cornstarch suspension, Milli-Q water (Millipore Corp., resistivity 18.2 M$\Omega$.cm), dyed with rhodamine B (Sigma-Aldrich) for enhancing the contrast at the interface, is injected into the HS cell as the inner displacing fluid at a controlled injection flow rate $q$. 
The growth of interfacial patterns is recorded using a DSLR camera (D5200, Nikon, Japan) with a spatial resolution of 1920$\times$1080 pixels (one-pixel area = $2.2 \times 10^{-3}$ cm$^2$) and a frame rate of 30~fps. The obtained stack of images is converted to grayscale format and analysed using the MATLAB@2021 image processing toolbox. The procedure for binarization of raw images is shown in Fig.~\ref{fig:hssem}(b).  {Snapshots of raw images corresponding to the temporal evolutions of interfacial patterns for displacing fluid flows lying in the low, intermediate and high ranges are shown in Supplementary Figs.~S1, S2 and S3 respectively.} A stress-controlled rheometer (Anton Paar, MCR 702) is used to perform rheological measurements in a parallel plate geometry (PP50) at different plate separations. 
 {Figure~\ref{fig:hssem}(c) shows the plots of the measured shear rate dependent viscosities of CS suspensions, $\eta_{out}$, plotted versus applied rotational shear rates, $\dot{\gamma}$, for different cornstarch particle concentrations at a plate separation of 300 $\mathrm{\mu}$m.} Shear-thickening, an increase in viscosity of the fluid with increasing $\dot{\gamma}$ above a critical shear rate $\dot{\gamma}_c$~\cite{doi:10.1122/1.3696875}, is prominent in the CS suspensions prepared at high concentrations. We observe from the inset of Fig.~\ref{fig:hssem}(c) that $\dot{\gamma}_c$ decreases with increase in concentration of the CS suspension~\cite{doi:10.1122/1.3696875}. We note that the observed decrease in the viscosities of 40 wt.\% and 42 wt.\% CS suspensions at very high shear rates arises due to the slippage of the dense CS suspensions at the stainless steel rheometer plates. All the displacement experiments and rheological measurements are performed at room temperature (25$^{\circ}$C).

\subsection{Calculations}
We estimate the shear rate $\dot\gamma$ imposed by water (the inner fluid) on the cornstarch (CS) suspension (outer displaced fluid) during displacement of the latter in the Hele-Shaw cell. The shear rate imposed by a propagating finger-tip is computed using $\dot\gamma$= 2$U/b$~\cite{Nagastu}, where $U$ is the characteristic radial propagation velocity of the interfacial finger-tips and $b$ is the gap of the Hele-Shaw cell. The finger-tip velocity, $U$, was estimated by tracking the temporal propagation of finger-tips at the inner interface using video imaging. Since each finger-tip in a pattern experiences a different local shear rate, an average over multiple finger-tips is calculated to estimate $\dot\gamma$ values for different injection flow rate experiments (Supplementary Fig.~S4). The estimated values of $\dot\gamma$ vary from $\num{6.01} \pm 0.95~\mathrm{s^{-1}}$ to $\num{230.35}\pm 52.72~\mathrm{s^{-1}}$ for the displacement of a dense 40 wt.\% CS suspension and lie above the critical shear rate $\dot{\gamma}_c = 1.33~\mathrm{s^{-1}}$ required for the onset of shear-thickening  behaviour (shaded region in Fig.~\ref{fig:hssem}(c)).

 \begin{figure}[!ht]
		\includegraphics[width=0.8\textwidth]{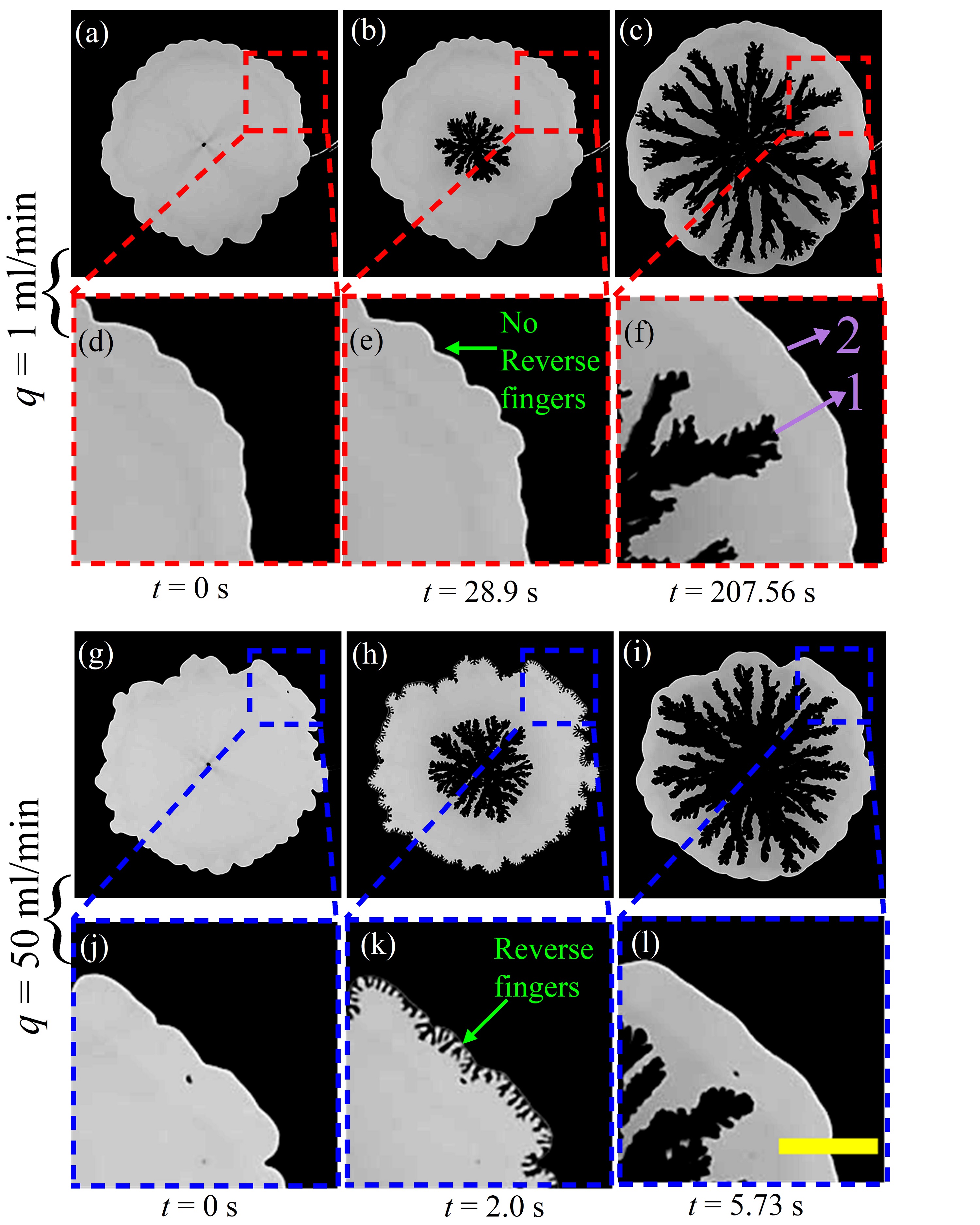}
		\centering
		\caption{\label{patterns}{\bf Temporal evolution of interfacial patterns at two different injection flow rates.} \textbf{(a-c)} Patterns in grayscale formed during the displacement of a 40 wt.\% cornstarch suspension (displaced outer fluid) by water (inner displacing fluid) at injection flow rate $q$ = 1 ml/min at different times of their growth. \textbf{(d-f)} Zoomed images of the patterns within the red-coloured boxes in (a-c) are displayed. The inner interface between water and cornstarch suspension, and the outer interface between cornstarch suspension and air (outermost fluid) are indicated by numbers 1 and 2 respectively in (f). \textbf{(g-i)} Patterns in grayscale for injection flow rate $q=$ 50 ml/min. \textbf{(j-l)} Zoomed images of the patterns within the blue-coloured boxes in (g-i) are displayed. The scale bar is 3 cm. The HS cell gap is 170 $\mathrm{\mu}$m.}
	\end{figure}

\section{Results \& Discussions}
 Figures~\ref{patterns}(a-c) display grayscale images showing the temporal evolution of interfacial patterns during the displacement of an aqueous 40 wt.\% cornstarch (CS) suspension (displaced outer fluid) by water (inner displacing fluid) at a low injection flow rate $q$ = 1 ml/min \href{https://youtube.com/shorts/0aAr3KEPGc4}{ {(Supplementary Video 1)}}  for a HS cell gap $b$ = 170 $\mathrm{\mu}$m. The magnified images of the interfacial regions enclosed in red boxes in Figs.~\ref{patterns}(a-c) are shown in Figs.~\ref{patterns}(d-f). In this work, we simultaneously explore the growth of two interfaces: the inner interface between water and the CS suspension (inner interface labelled as 1 in Fig.~\ref{patterns}(f)) and the outer interface between CS suspension and air (outer interface labelled as 2 in Fig.~\ref{patterns}(f)). The growth of the inner interface due to outward displacement of the CS suspension involves the appearance of fingers undergoing multiple tip-splitting events, as seen in Figs.~\ref{patterns}(b-c) and \href{https://youtu.be/o5GoqPFtniQ}{ {(Supplementary Video 2)}}.  {These fingers at the inner interface also display coalescence~\cite{shokri,PALAK2021127405}. Since water, the displacing fluid in our experiments, is less viscous than the outer cornstarch suspension, the inner interface in our experiments tends to destabilise as soon as injection of the displacing fluid is initiated. Furthermore, we note from Supplementary Figs.~S1, S2 and S3 that the morphologies of the inner interfacial patterns are insensitive to the injection flow rate of the displacing water.} Figures~\ref{patterns}(g-i) show grayscale images of the temporal evolution of interfacial patterns during the displacement of a 40 wt.\% CS suspension by water at a very high injection flow rate $q$ = 50 ml/min. Interestingly, we note the transient withdrawal of the CS suspension at the outer interface during its displacement at high injection flow rates (Fig.~\ref{patterns}(h), \href{https://youtube.com/shorts/iTZsj01IqM0?feature=share}{ {(Supplementary Video 3)}}). This withdrawal process results in the invasion of air (outermost fluid) into the cornstarch suspension and the development of reverse fingers at the outer interface (Fig.~\ref{patterns}(h)) between the CS suspension and air. Magnified images of the regions enclosed in blue boxes in Figs.~\ref{patterns}(g-i) are displayed in Figs.~\ref{patterns}(j-l). The reverse fingers appear for a very short time, and the outer interface eventually becomes smooth at later stages regardless of the applied injection flow rate (Fig.~2(l)).  {
 One of the key aims of this paper is to study the emergence and disappearance of the observed transient reverse fingers and correlate their growth with the evolution of the inner interface. }

 \begin{figure}[!b]
		\includegraphics[width= 6.6in ]{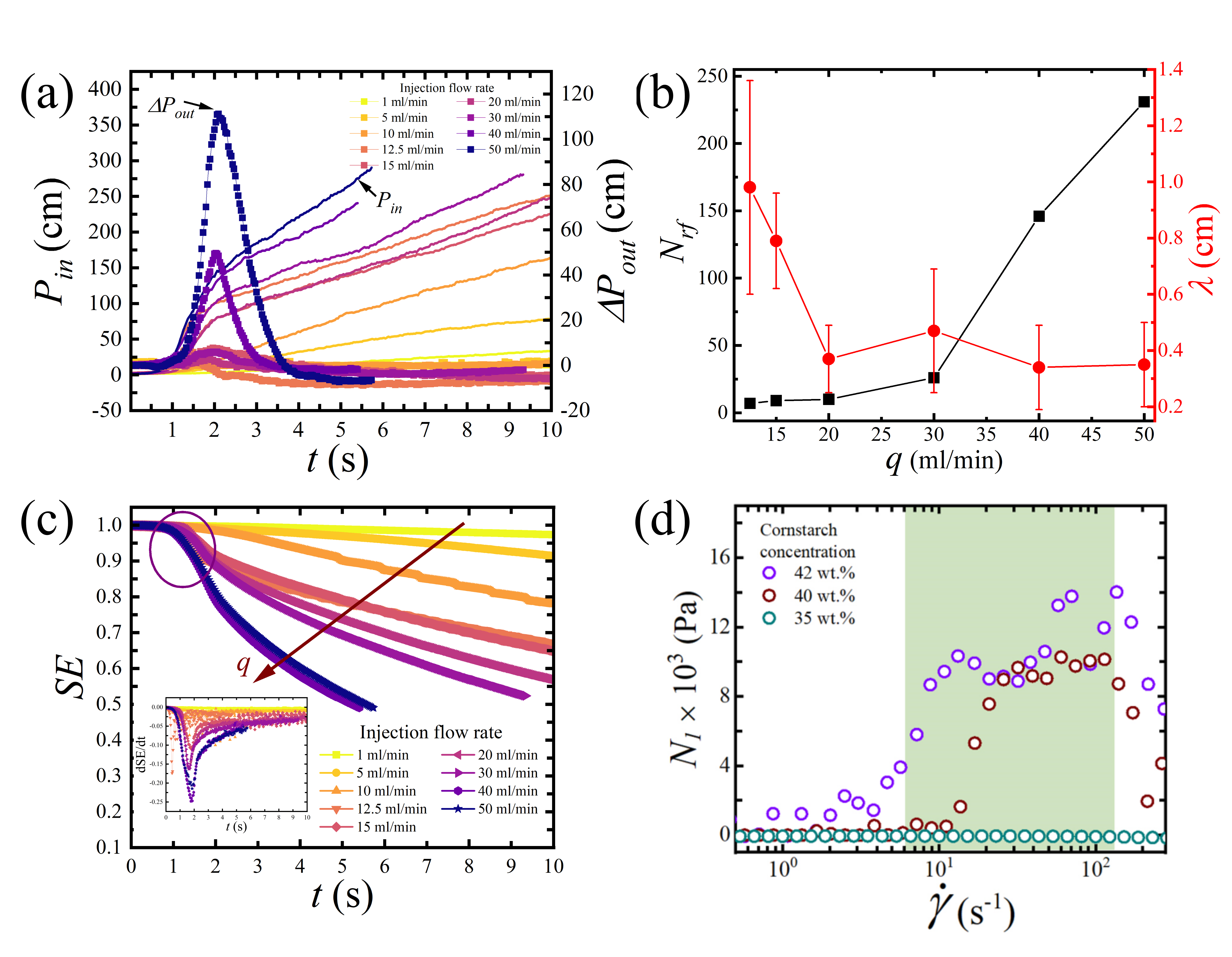}
		\centering
		\caption{\label{perimeter}{\bf Characterisation of the inner interface between water and cornstarch suspension and the outer interface between cornstarch suspension and air with injection flow rate $q$ as a control parameter.} \textbf{(a)} Perimeters of inner interfaces $P_{in}$ (solid lines) and outer interfaces $\Delta P_{out} = P_{out}(t) - P_{out}(0)$ (filled symbols connected by solid lines) vs. time $t$ where $P_{out}(t)$ and $P_{out}(0)$ are perimeters at times $t$ and $t$ = 0 s respectively at various injection flow rates, with $t$ = 0 s corresponding to the time of injection of the inner displacing fluid (water). \textbf{(b)} Number of reverse fingers $N_{rf}$ (\blsquare) and average reverse finger spacing $\lambda$ (\rcircle) as a function of injection flow rate $q$. \textbf{(c)} Sweep efficiency $SE$ vs. time $t$ for different injection flow rates of the inner fluid. A purple ellipse highlights the sharp decrease in $SE$ due to the generation of reverse fingers. Inset shows $dSE/dt$ vs. time. \textbf{(d)} First normal stress difference $N_1$ vs. applied shear rate for cornstarch suspensions of different concentrations measured in a stress-controlled rheometer at a plate separation of 300 $\mathrm{\mu}$m in a parallel plate experimental geometry. The shear rates imposed by the inner displacing fluid in the HS cell experiments at various injection flow rates, estimated as described in section 2.1, lie in the region highlighted in green.}
	\end{figure}

 We next quantify the morphologies and growth of the inner and outer interfaces (Fig.~\ref{perimeter}). It is seen from Fig.~\ref{perimeter}(a) that the perimeter of the inner interface, $P_{in}$, increases monotonically with time (Fig.~\ref{perimeter}(a)) showing very rapid initial growth, followed by a significant slowing down at later times. The perimeter of the outer interface is defined as $\Delta P_{out} = P_{out}(t) - P_{out}(0)$ where $P_{out}(t)$ and $P_{out}(0)$ are perimeters of the outer interface at times $t$ and $t$ = 0 s respectively, with $t$ = 0 s corresponding to the time of injection of water (inner displacing fluid). While $\Delta P_{out}$ does not change appreciably for low injection flow rates of water, we note that it shows a non-monotonic variation with time for high injection flow rates. The initial rapid increase in $\Delta P_{out}$ for high injection flow rates is due to the formation of reverse fingers at the outer interface between the cornstarch suspension and air. After reaching a maximum, the subsequent decrease in $\Delta P_{out}$ is attributed to the fading of reverse fingers at later times. We note from Fig.~\ref{perimeter}(a) that changes in the slopes of $P_{in}$ and $\Delta P_{out}$ occur at almost the same time, thereby indicating a strong correlation in the growth kinetics of the inner and outer interfaces. The variations in the time derivatives of $P_{in}$ and $\Delta P_{out}$, $i.e.$ in the growth rates of the inner and outer interfaces, further confirm this correlation (Supplementary Fig.~S5). Such close correlation between the dynamics of the interfaces 1 and 2 (Fig.~\ref{patterns}(f)) suggests that the stresses generated within the dense cornstarch suspension during its displacement by the inner displacing
 fluid are transmitted to the outer interface between the CS suspension and air.

  {We first note the emergence of reverse fingers at an inner fluid injection flow rate between 10 ml/min and 12.5 ml/min (Supplementary Figs.~S1, S2), from which we estimate a critical shear rate $\dot{\gamma}_{c^{\prime}}$ for the appearance of reverse fingers, with 53.42 $s^{-1} < \dot{\gamma}_{c^{\prime}} <$ 87.44 $s^{-1}$ during the displacement of 40 wt.\% CS suspensions.} The global features of the outer interface are next quantified by computing the number of reverse fingers, $N_{rf}$, and the average spacing between reverse fingers, $\lambda$, for different injection flow rates of water (12.5 ml/min $\leq q \leq$ 50 ml/min, Fig.~\ref{perimeter}(b)). The number of reverse fingers, $N_{rf}$ is estimated by identifying all the tips of the reverse fingering pattern at the outer interface when the transient reverse fingers reach their maximum lengths. We see that the $N_{rf}$ increases with increase in injection flow rate (Fig.~\ref{perimeter}(b)). The average spacing between the reverse fingers is estimated as $\lambda = {<\sqrt{(r_1)^2 + (r_2)^2 + 2 r_1 r_2 \cos(\theta_1 - \theta_2)}>}_{N_{rf}-1}$, where $r$ and $\theta$ are polar coordinates corresponding to the tips of the reverse fingers and $<...>_{N_{rf}-1}$ denotes an average over the estimated spacings between all the adjacent reverse finger-tips (Fig.~\ref{perimeter}(b)). The average reverse finger spacing $\lambda$  does not show any significant dependence on the injection flow rates explored in our experiments. Further details about the calculations of $N_{rf}$ and $\lambda$ are provided in Supplementary Fig.~S6. 

Sweep efficiency ($SE$) is a non-dimensional parameter often used to determine how effectively one fluid displaces another~\cite{PALAK2021127405,shokri}. By following the protocols adopted in our previous report~\cite{PALAK2021127405},  {we estimate $SE$ of the displaced cornstarch suspension by computing the ratio of the area occupied by the outer displaced fluid ( {white} area in Fig.~\ref{fig:hssem}(a)) to the total area occupied by the inner displacing and outer displaced fluids (sum of the black and white areas in Fig.~\ref{fig:hssem}(a)).} Figure~\ref{perimeter}(c) shows the variation in sweep efficiency with time for all the injection flow rates $q$ used in our experiments. Before the onset of pattern growth at the inner interface at the earliest times, we observe that $SE$ is close to unity for all $q$ values. This is followed by a decrease in $SE$ for the higher flow rates as the interfacial patterns evolve at later times. The initial sharp drops of $SE$ values occur at the same times when reverse fingers are observed (highlighted by a purple ellipse in Fig.~\ref{perimeter}(c)). This is also evident from the slopes of the $SE$ vs. $t$, $dSE/dt$ shown in the inset of Fig.~\ref{perimeter}(c), thereby confirming that the presence of reverse fingers significantly affects sweep efficiency during displacement of the highly sheared cornstarch suspension.

It has been predicted in non-linear simulations that non-zero normal stresses in a viscoelastic fluid lead to significantly higher stress asymmetries along the flow direction when compared to the normal direction~\cite{shokri}. Figure~\ref{perimeter}(d) displays the rapid increase in the first normal stress difference $N_1$ with shear rate for the higher cornstarch suspension concentrations. The observed decrease in $N_1$ at very high shear rates can be attributed to slip~\cite{wallslip2,wallslip,macosko1994rheology} between the extremely dense CS suspension and the rheometer plates due to the imposition of large tangential strains. It is important to note that such slippage-induced decreases in $N_1$ (Fig.~\ref{perimeter}(d)) and $\eta_{out}$ (Fig.~\ref{fig:hssem}(c)) are observed at comparable shear rates. However, as confirmed from the temporal variations of the perimeters ($P_{in}$ and $\Delta P_{out}$) of the interfaces (Fig.~\ref{perimeter}(a)),  we do not observe any intermittent changes in growth profiles of the inner and outer interfaces in the HS cell. We therefore expect minimal or no slippage between the suspension and the HS glass plates in our displacement experiments. It was reported earlier that the sign of $N_1$ depends on the details of the particle-particle interactions in shear-thickened suspensions of colloidal silica~\cite{normal} and granular cornstarch~\cite{normal1,PhysRevResearch.4.033062}. While negative values of $N_1$ represent hydrodynamic or lubrication effects, positive values of $N_1$ reflect the dominant influence of inter-particle friction~\cite{normal}. In our experiments, the measured values of $N_1$ (Fig.~\ref{perimeter}(d)) at the imposed shear rates are always positive, indicating that for high concentrations and injection flow rates of the displacing fluid, the displaced cornstarch suspension is in the discontinuous shear thickening (DST) regime, with the friction between particles assembled in force chain networks~\cite{frictional} determining its rheology.

 {Large-scale intermittent formation of anisotropic force chain networks~\cite{doi:10.1073/pnas.2203795119,Jamali,CST,Majmudar2005} that are sustained by frictional contacts between particles guide the anisotropic transmission of stresses in the suspension~\cite{brown2012role}. When subjected to shear rate above a critical value $\dot{\gamma}_{c^{\prime}}$, an aqueous cornstarch suspension dilates momentarily~\cite{maharjan}. This causes the CS particles to penetrate the suspension-air interface~\cite{brown2012role, maharjan}. The surface roughness that is generated as a consequence drives a restoring force which results in the migration of air into the CS suspension and the generation of the observed reverse fingers. We note here that the anisotropic stresses propagating through the intermittent and spatially localized force chains also affect pattern growth at the inner interface, with the velocity of the inner perimeter decreasing abruptly when the reverse fingers start fading away, as shown in Supplementary Fig.~S5(b).} 
 \begin{figure}[htbp]
    \centering
    \includegraphics[width=1\textwidth]{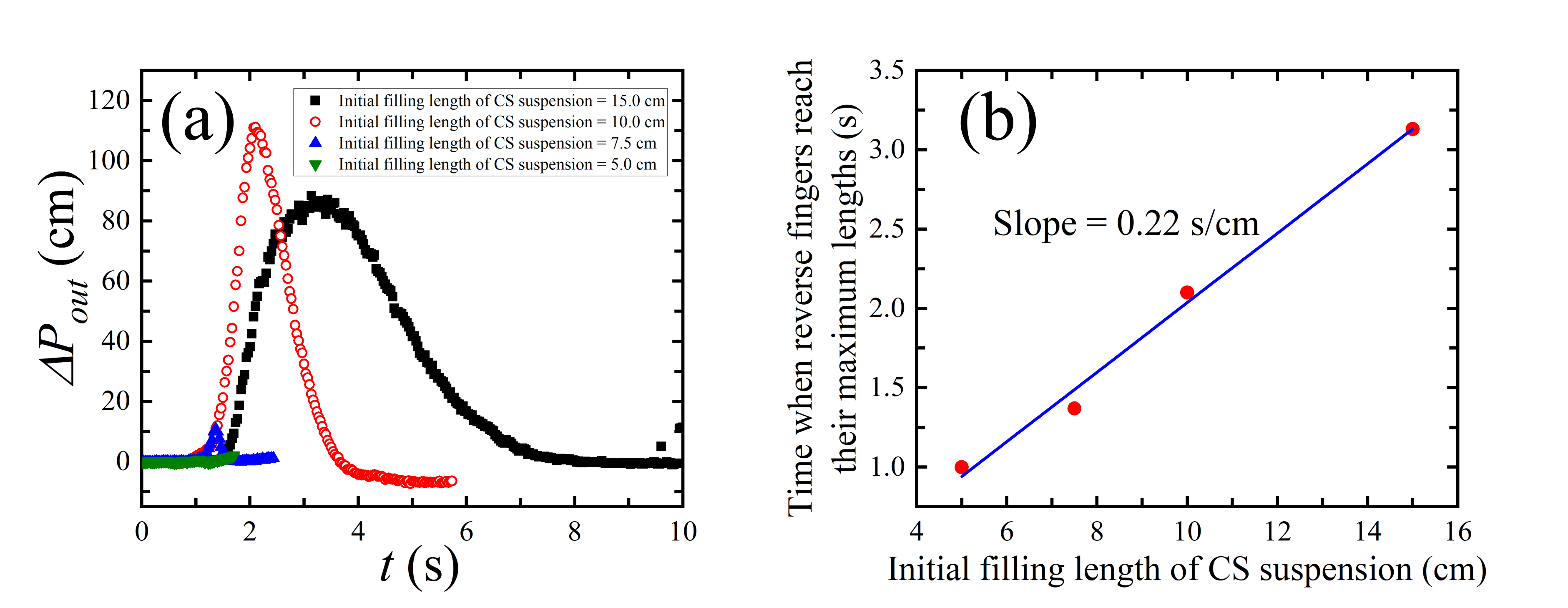}
    \caption{\textbf{Effect of initial filling length of cornstarch (CS) suspension on the formation of reverse fingers at the outer interface} (a) Perimeter of the outer interface ($\Delta P_{out}$ = $P(t)$ – $P(0)$) vs. time $t$ for different CS suspension filling lengths. (b) The time corresponding to maximum transient withdrawal of the CS suspension and appearance of longest reverse fingers is plotted against the initial filling length of the suspension. The data is fitted to a straight line and stress propagation velocity is extracted by computing the reciprocal of the slope. The longer durations elapsed for the appearance of reverse fingers for longer initial filling lengths is consistent with our claim of propagation of stresses via force chain networks.}
    \label{fig:varFilling}
\end{figure}

A recent study reported a characteristic time of a few seconds for the applied stresses to propagate across highly sheared cornstarch suspensions undergoing discontinuous shear-thickening~\cite{maharjan}. These stresses are anisotropically transmitted via particle-particle contacts in sheared force networks that form and break under large shear rates.  {In our displacement experiments involving 40 wt.\% CS suspensions, the maximum growth of transient reverse fingers occurs at t $\approx$ 2 seconds for high injection flow rates (as seen from the peak positions of $\Delta P_{out}$ in Fig.~\ref{perimeter}(a)) and coincides approximately with the expected time interval for the propagation of stresses through force networks in highly sheared dense cornstarch suspension as determined in an earlier study~\cite{maharjan}. The subsequent rearrangement of these fragile force networks under very high shears at later times results in the disappearance of the observed reverse fingers. To support our argument that transient suspension withdrawal and emergence of reverse fingers at the outer interface occur due to the build-up of suspension-spanning anisotropic stresses under very high shear rates, we perform a set of experiments to determine the approximate values of the stress propagation times and velocities in highly sheared dense cornstarch suspensions. We change the initial filling length of the CS suspension by injecting different suspension volumes into the HS cell and perform miscible displacement experiments by injecting water at $q$ = 50 ml/min. We observe the transient withdrawal of cornstarch suspension irrespective of the initial suspension filling length (Supplementary Fig.~S7, Fig.~\ref{fig:varFilling}(a)). Grayscale images of the pattern corresponding to the time at which reverse fingers reach their maximum sizes for different CS suspension filling lengths are shown in Supplementary Fig. S7. We note that the time required for reverse fingers to reach their maximum lengths increases linearly with the initial filling length of the cornstarch suspension as seen in Fig.~\ref{fig:varFilling}(b). This linear variation allows us to determine a propagation velocity of stresses, $\sim$ 4.52 cm/s, across the cornstarch suspension under these conditions.} 
\begin{figure}[!b]
	 	\includegraphics[width= 6.0in ]{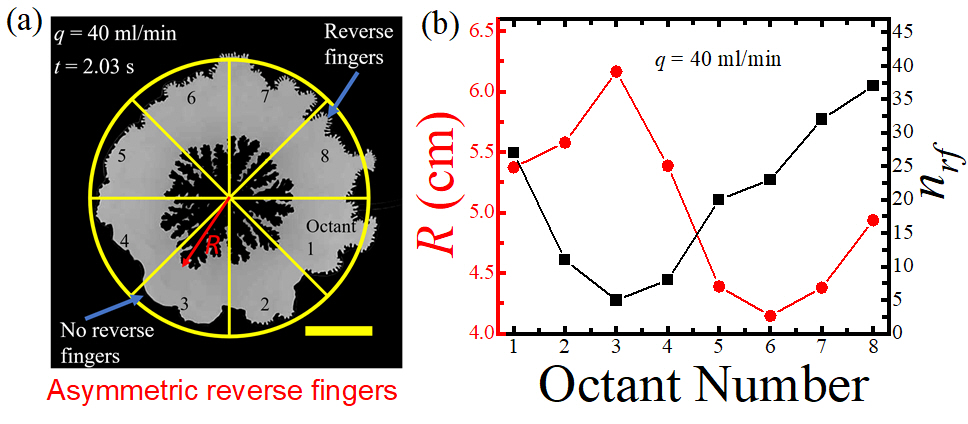}
	 	\centering
	 	\caption{\label{asymmetric}{\bf Asymmetry in the generation of reverse fingers at the outer interface between cornstarch suspension and air.} \textbf{(a)} Interfacial pattern obtained during the radial displacement of a cornstarch suspension (40 wt.\%) by water at $q$ = 40 ml/min for HS cell gap $b$ = 170 $\mathrm{\mu}$m. The scale bar is 5 cm. Formation of reverse fingers is not axisymmetric in the different interfacial sections. Yellow lines divide the pattern into octants, which are labelled from numbers 1 to 8. $R$ is the length of the longest finger of the inner pattern between the cornstarch suspension and water and $n_{rf}$ is the number of reverse fingers in an octant. \textbf{(b)} The longest finger length $R$ and the number of reverse fingers  $n_{rf}$ in each octant of the pattern displayed in (a) vs. octant number.} 
	 \end{figure}
  
As seen in Fig.~\ref{asymmetric}(a), the occurrence of reverse fingers is not always axisymmetric at the outer interface between the cornstarch suspension and air. We further analyse the reverse fingering patterns for high injection flow rates by dividing each image into eight octants, with each octant labelled by a unique number between 1 and 8 as shown in Fig.~\ref{asymmetric}(a). For each octant, the number of reverse fingers, $n_{rf}$, at the outer interface and the longest finger length, $R$, of the inner pattern are estimated. The longest finger length in the inner pattern, $R$, varies appreciably in the different octants and is approximately inversely proportional to $n_{rf}$ as seen in Fig.~\ref{asymmetric}(b). This indicates that growth of the inner pattern is comparatively slower in the interfacial sections having more reverse fingers $n_{rf}$. While we may expect a decrease in the finger spacing $\lambda$ with an increasing number of equally distributed reverse fingers, we note that the constant value of $\lambda$ with changing $q$ as seen in Fig.~\ref{perimeter}(b) also indirectly indicates that the reverse fingers are not necessarily axisymmetric (Fig.~\ref{asymmetric}(a)). It is reasonable to conclude that the anisotropic build-up of normal stresses~\cite{Majmudar2005,doi:10.1073/pnas.2203795119,CST} in the displaced cornstarch suspension causes non-uniform drag forces across the sample. These drag forces lead to the observed slower growth of the propagating finger-tips of the inner pattern in certain sections, an effective withdrawal of the cornstarch suspension, and the generation of pronounced reverse fingers at the outer interface.

	\begin{figure}[!t]
	 	\includegraphics[width= 6.0in ]{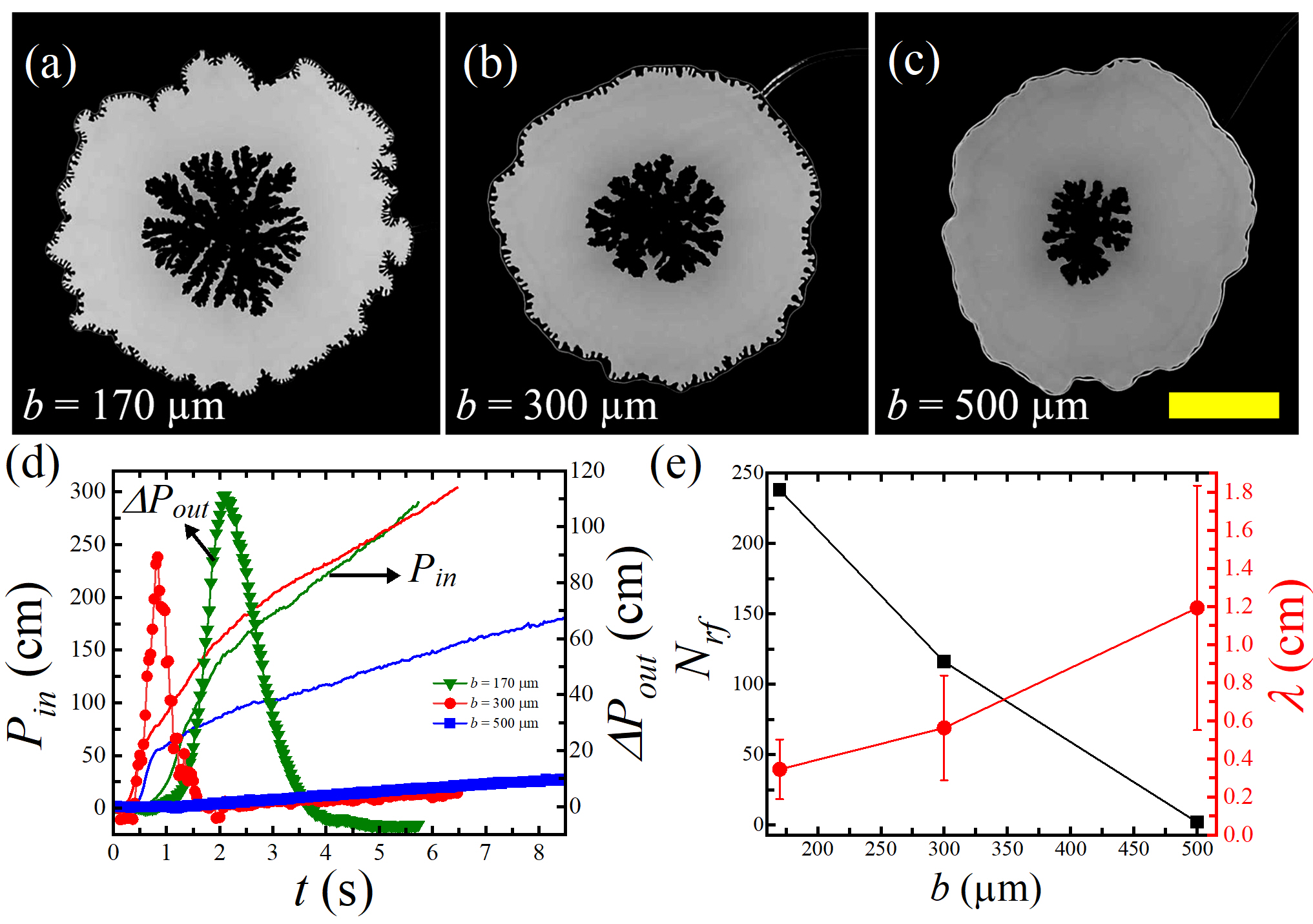}
	 	\centering
	 	\caption{\label{gap}{\bf Characterisation of the inner interface between water and cornstarch suspension, and the outer interface between cornstarch suspension and air while increasing gap $b$ of the Hele-Shaw cell.} Interfacial patterns obtained during the displacement of a 40 wt.\% cornstarch suspension by water at $q$ = 50 ml/min for different gaps $b$ of the Hele-Shaw (HS) cell: \textbf{(a)} $b$ = 170 $\mathrm{\mu m}$, \textbf{(b)} $b$ = 300 $\mathrm{\mu m}$ and \textbf{(c)} $b$ = 500 $\mathrm{\mu m}$. The scale bar is 5 cm. \textbf{(d)} Perimeters of inner interfaces $P_{in}$ (solid lines) and outer interfaces $\Delta P_{out}$ (filled symbols connected by solid lines) for different gaps $b$. \textbf{(e)} Number of reverse fingers $N_{rf}$ (\blsquare) and average reverse finger spacing $\lambda$ (\rcircle) at the outer interface as a function of $b$.}
	 \end{figure}

 We next investigate the effect of confinement on the reverse fingering patterns when a dense cornstarch suspension (40 wt.\%) is displaced by water at a fixed injection flow rate $q$ = 50 ml/min in a Hele-Shaw cell with gaps $b$ varying between 170 $\mathrm{\mu}$m to 500 $\mathrm{\mu}$m (Figs.~\ref{gap}(a-c)). We see the formation of reverse fingers for the lower HS cell gaps (HS cell gap $b$ = 170 and 300 $\mathrm{\mu}$m, Figs.~\ref{gap}(a-b)), but observe only a slight withdrawal of the outer interface and the absence of reverse finger formation at $b$ = 500 $\mathrm{\mu}$m (Fig.~\ref{gap}(c)). When $b$ is increased to 800 $\mathrm{\mu}$m, we note that water, the inner fluid, spreads over the cornstarch suspension rather than displacing it \href{https://youtube.com/shorts/vSyPHpdg2A0}{ {(Supplementary Video 4)}}.  {The spreading occurs because of the inability of the injected water to completely displace the CS suspension effectively across the rather broad HS cell gap, as illustrated in Supplementary Fig. S8}. We next quantify the effects of confinement on the onset and growth of the patterns at the inner and outer interfaces by estimating the pattern perimeters $P_{in}$ and $\Delta P_{out}$, number of reverse fingers $N_{rf}$ and average reverse finger spacing $\lambda$ for different values of $b$. As reported by us for pattern formation at a fixed HS cell gap and different injection flow rates (Fig.~\ref{perimeter}(a)), $P_{in}$ increases monotonically with time and shows a change in slope at intermediate times while $\Delta P_{out}$ shows a non-monotonic time-dependence (Fig.~\ref{gap}(d)) for the smaller HS cell gaps. Decrease in the peak values of $\Delta P_{out}$ with increase in $b$ (Fig.~\ref{gap}(d)) is a consequence of a decrease in the number of reverse fingers $N_{rf}$ (Fig.~\ref{gap}(e)) with the removal of confinement. Simultaneously, $\lambda$ is seen to increase with $b$ (Fig.~\ref{gap}(e)). As demonstrated earlier, the formation of reverse fingers is sensitively dependent on the first normal stress difference $N_1$. Our rheometric measurements reveal that $N_1$ decreases steadily with increasing gap thickness of the rheometer plates (Supplementary Fig.~S9). This is consistent with previous work that highlighted the increasingly strong shear-thickening rheology of cornstarch suspensions with decreasing plate separations~\cite{doi:10.1122/1.3696875}. The large values of $N_1$ at low rheometer plate separations indicate that confined geometries are necessary for the generation of reverse fingers. 
 
 \begin{figure}[ht!]
    \centering
    \includegraphics[width=0.5\textwidth]{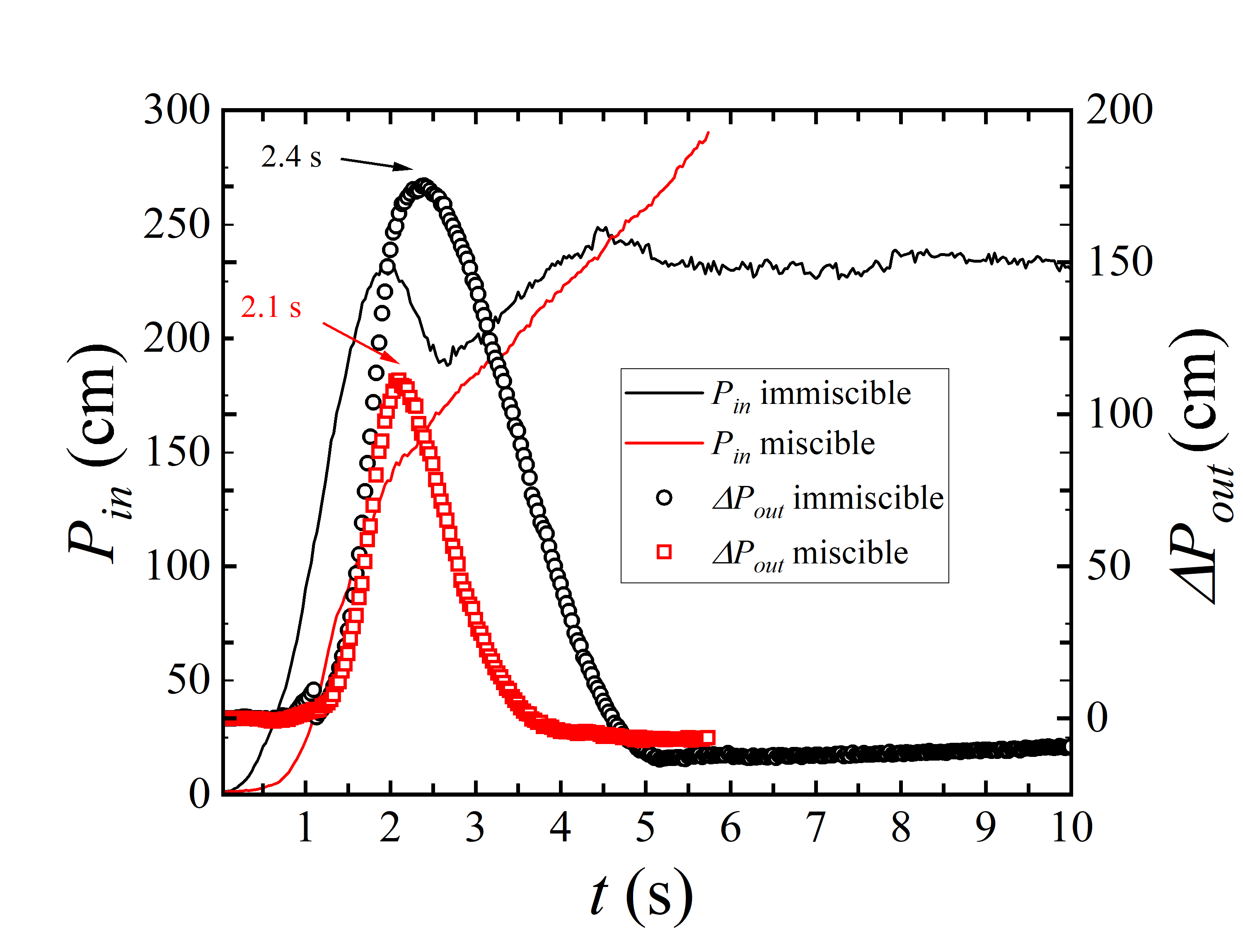}
    \caption{\textbf{Effect of miscibility on the formation of reverse fingers at the outer interface} Perimeter of the outer interface ($\Delta P_{out}$ = $P(t)$ – $P(0)$) vs. time $t$ for miscible and immiscible displacements. We observe that reverse fingers reach their maximum lengths sightly later for immiscible displacements.}
    \label{fig:6}
\end{figure}

 {The anisotropic stresses generated in dense cornstarch (CS) suspensions are generated at sufficiently high shear rates and are believed to be supported by the intermittent 
 formation of fragile force chain networks~\cite{doi:10.1073/pnas.2203795119,Jamali,CST,Majmudar2005}. In our experiments reported so far, we have displaced confined CS suspensions with miscible water at high injection flow rates and have studied the growth of reverse fingers. To confirm our hypothesis that reverse fingers originate due to the build-up of system-spanning anisotropic stresses in highly sheared cornstarch suspensions, we next displace a 40 wt.\% CS suspension with an immiscible mineral oil at an injection flow rate of 50 ml/min and observe the temporal variation of the outer interface. Despite the immiscibility of the displacing fluid, we observe the formation of reverse fingers at the outer interface between the cornstarch suspension and atmospheric air (Supplementary Figs.~S10(a1-a6)). However, we simultaneously note that finger propagation and the growth of the inner interface are very sensitive to the miscibility of the fluid pair (Supplementary Fig.~S10). Interfacial patterns that form at the inner interface depend on the viscosity contrast between the fluid
 pair~\cite{saffman1958penetration,bischofberger2014fingering,PALAK2021127405}. In immiscible displacements, the inner pattern growth is further stabilized by non-zero interfacial tension~\cite{chen1989growth} which is absent in miscible displacement experiments. In our immiscible displacement experiments (Supplementary Figs.~S10(a1-a6)), we observe the generation of multiple fingers at the inner interface that resemble dendritic fractures. Dendritic fractures were reported in a recent article~\cite{ozturk2020flow} when a cornstarch suspension was displaced by immiscible air. In this study, the authors focused only on the inner interface and reported a transition from viscous fingering to dendritic fracturing followed by system-wide fracturing with increasing particle volume fraction or injection pressure. They showed that dendritic fractures form when the displaced cornstarch suspension momentarily goes to a discontinuous shear-thickening solid-like state. The suspension relaxes back to the fluid state immediately after the passage of the immiscible invasion front. This is accompanied by an increase in the widths of the fingers at the inner interface.\\
In our case, the cornstarch suspension lies in a discontinuous shear thickening (DST) regime. Similar to the results discussed above~\cite{ozturk2020flow}, the CS suspension in our experiments temporarily transforms to a solid state due to the high shears applied by the displacing fluid. The increase in perimeter of the outer pattern, $\Delta P_{out}$ with time at the initial stages (Fig.~\ref{fig:6}) arises due to the growth of reverse fingers at the outer interface due to the displacement of the discontinuous shear thickening CS suspension. After peaking, $\Delta P_{out}$ decreases at later times due to the fading away of reverse fingers. The increase in the width of the propagating fingers at the inner interface at later times as the reverse fingers start fading away is evident from the decrease in $P_{in}$ values for immiscible displacements, as shown in Fig.~\ref{fig:6}. This suggests that the fluid relaxes back to a liquid-like state after a few seconds, as suggested in~\cite{ozturk2020flow}. The simultaneous increase in the width of the fingers at the inner interface and fading of the reverse fingers at the outer interface as seen in immiscible displacement experiments is yet another indication that the propagation of stresses through force chain networks in highly-sheared CS suspensions is responsible for the evolution of reverse fingers.  We note that reverse fingers reach their maximum lengths slightly later in immiscible displacement experiments (red in Fig.~\ref{fig:6}), when compared to the ones involving miscible displacements (black in Fig.~\ref{fig:6}). We note from Supplementary Fig.~S10 that the average reverse finger spacing $\lambda$ is also marginally longer for immiscible displacements when compared with miscible displacements ($0.41 \pm 0.19$ cm vs $0.35 \pm 0.15$ cm in the two cases). Moreover, we note that the average velocity of the longest propagating finger in the case of immiscible displacement ($\bar{U}$ = 0.66 cm/s) is less than its miscible counterpart ($\bar{U}$ = 1.92 cm/s).}
	  \begin{figure}[!t]
	 	\includegraphics[width= 5.0in ]{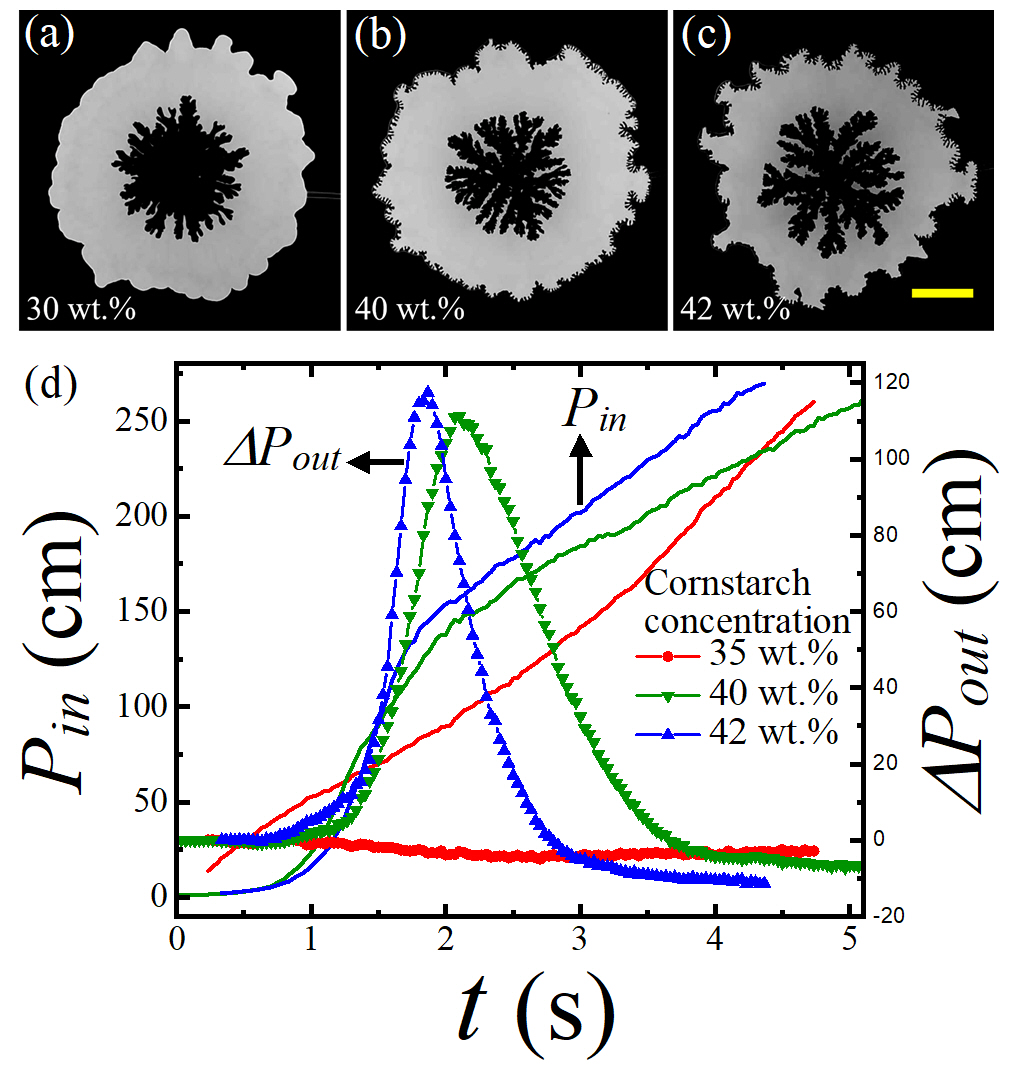}
	 	\centering
	 	\caption{\label{Varconc}{\bf Characterisation of the inner interface between water and cornstarch (CS) suspension, and the outer interface between cornstarch suspension and air while increasing the concentration of the CS suspension.} Interfacial patterns formed by the displacement of cornstarch suspensions of various concentrations: (a) 35 wt.\% (b) 40 wt.\% (c) 42 wt.\% at time $t$ = 2 s after injection of water at $q$ = 50 ml/min in the HS cell of gap $b$ = 170 $\mathrm{\mu}$m. The scale bar is 5 cm. (d) Perimeters of the inner interfaces $P_{in}$ (solid lines) and outer interfaces $\Delta P_{out}$ (filled symbols connected by solid lines) vs. time $t$ for the above cornstarch suspensions.}
	 \end{figure}
  
 We therefore conclude that the formation of reverse fingers at the outer interface requires the build-up of large normal stresses in the cornstarch suspension and can be achieved either by increasing the injection flow rate of the inner displacing fluid or by decreasing the gap of the HS cell.  {Furthermore, we have established systematically that the occurrence of reverse fingers is independent of miscibility of the fluid pair as long as the shear rate applied by the displacing fluid is high enough that the displaced CS suspension lies in the DST regime}. The important role of the first normal stress difference in the generation of reverse fingers is further confirmed by increasing the concentration and therefore the elasticity~\cite{PALAK2021127405} of the displaced cornstarch suspension. Figures~\ref{Varconc}(a-c) display grayscale images of the interfacial patterns formed when water injected at an injection flow rate $q$ = 50 ml/min displaces cornstarch suspensions of different concentrations in a Hele-Shaw cell of gap $b$ = 170 $\mathrm{\mu}$m.  While reverse fingering is observed when cornstarch suspensions of higher concentrations are displaced by water, it is absent when a cornstarch suspension of a lower concentration (35 wt.\%) is displaced at the same injection flow rate. The higher peak value of the outer perimeter $\Delta P_{out}$ and its non-monotonic evolution with time at the highest concentration (42 wt.\%; Fig.~\ref{Varconc}(d)) indicates the enhanced formation of reverse fingers at the outer interface. The variations in the time derivatives of $P_{in}$ and $\Delta P_{out}$ (Supplementary Figs.~S11(a-b)), $i.e.$ in the growth rates of the inner and outer interfaces, indicate correlation in the growth kinetics of the inner and outer interfaces. We conclude by noting that the time-evolutions of $P_{in}$ and $\Delta P_{out}$ qualitatively follow the same trend regardless of whether cornstarch suspensions of increasing concentrations are displaced at a fixed injection flow rate (Fig.~\ref{Varconc}(d)) or a cornstarch suspension of fixed concentration is displaced at increasing injection flow rates (Fig.~\ref{perimeter}(a)).

\section{Conclusions}
Dense aqueous granular cornstarch (CS) suspensions display shear-thinning and shear-thickening flows when externally applied shear stresses and particle concentrations are varied appropriately~\cite{peters2016direct}. It is now well-known that hydrodynamic lubrication forces and the formation of anisotropic force chain networks govern the unique rheology of cornstarch suspensions~\cite{wagner2009shear,peters2016direct,doi:10.1073/pnas.2203795119,CST,Jamali}. The spatially anisotropic stresses arising from the formation of force chain networks~\cite{Majmudar2005, doi:10.1073/pnas.2203795119,Jamali} at large shear rates should influence the displacement efficiency~\cite{shokri} of a dense shear-thickening cornstarch suspension in a Hele-Shaw (HS) cell. In this work, we investigate miscible displacements of confined shear-thickening cornstarch (CS) suspensions (outer displaced fluid) when the shear rates imposed on the suspension are greater than the critical shear rate $\dot\gamma_c$ required for the onset of shear-thickening. Displacement of dense CS suspensions at large shear rates are achieved by systematically varying injection flow rate of the inner displacing fluid (water) and gap of the HS cell. We note that previous literature focussed exclusively on the propagation of the inner fluid front at the inner fluid-fluid interface~\cite{saffman1958penetration,Homsy,pinilla2021,ozturk2020flow,PALAK2021127405,PALAK2022100047,Linder2000Viscous,Zhang2021Structures}. Besides monitoring the generation of viscous fingers at the inner interface between water and the CS suspension, we report here an unexpected growth of transient reverse fingers at the outer interface between the CS suspension and air (outermost fluid) at sufficiently high injection flow rates. Our observation of an inverse relation between the number of reverse fingers at the outer interface and the rate of growth of the inner pattern establishes the presence of a strong correlation between the growth kinetics of the two interfaces. Our rheometric measurements of large positive values of the first normal stress difference in dense cornstarch suspensions at large applied shear rates ($\dot{\gamma} \geq \dot{\gamma}_c$) indicate inter-particle frictional interactions~\cite{normal,frictional} and the generation of shear-induced anisotropic force chain networks~\cite{Majmudar2005,doi:10.1073/pnas.2203795119,Jamali}.  {These intermittent force chain networks direct the transmission of stresses across the suspension during its displacement. We propose that the dilation of the confined CS suspension under shear and the surface tension-driven restoring force~\cite{brown2012role,maharjan} at the outer interface between the CS suspension and atmospheric air drives the formation of the observed reverse fingers. The reverse fingers start fading immediately after propagation of the dilation front. We note that the growth of the inner interfacial patterns and the reverse fingers are closely correlated and estimate a propagation velocity of 4.52 cm/s for the anisotropic stresses transmitted in the highly sheared dense cornstarch suspension. We also demonstrate that the stress propagation times are in agreement with a previous estimate~\cite{maharjan} and report an approximate range for the critical shear rate required to generate reverse fingers.}

We also note the enhanced formation of reverse fingers for low gap widths of the HS cell and high concentrations of the CS suspensions. Since our rheology experiments clearly demonstrate that normal stresses in the CS suspension increase with decreasing gap of the HS cell~\cite{doi:10.1122/1.3696875} and increasing suspension concentration~\cite{normal}, our results verify the important role of anisotropic formation of stresses in pattern formation during the displacement of viscoelastic suspensions in confined geometries.  {Interestingly, the emergence of reverse fingers is also observed during the immiscible displacement of cornstarch suspension. A comparison of the features characterizing miscible and immiscible displacements further verify our hypothesis that the propagation of large stresses through the intermittent and spatially localized force chains are responsible for the formation of reverse fingers. The correlation of the growth profile of the inner interface with the emergence of reverse fingers at the outer interface for both miscible and immiscible displacement experiments is yet another indication that the formation of temporary force chain networks in highly sheared dense cornstarch suspension drive the generation of reverse fingers at the outer interface.}

 The magnitude of the normal stresses generated in viscoelastic fluids such as in emulsions~\cite{foamsn} and polymeric solutions~\cite{polym} are strongly dependent on their individual internal microstructures. In order to thoroughly investigate the relation between normal stresses, sample microstructures and morphologies of interfacial displacement patterns, it would be interesting to systematically perform displacement experiments with different materials and externally imposed shear profiles.
 
 In a significant advance to our previous work~\cite{PALAK2021127405} where we proposed different experimental protocols for controlling interfacial instabilities during the miscible displacement of shear-thinning cornstarch suspensions, we report here the first experimental observation of reverse fingers at the outer interface between a highly sheared CS suspension and air during miscible displacement of the former by water. Since displacement efficiency~\cite{shokri} depends on the morphologies of the inner and outer interfaces, we observe a sharp decrease in sweep efficiency due to the generation of reverse fingers at the outer interface. The role of shear-dependent rheology in the formation of interfacial patterns during the displacement of a more viscous fluid by a less viscous one is of fundamental and practical interest. Besides being fascinating from a fluid mechanics point of view, the understanding of interfacial instabilities can be useful in many areas such as hydrology~\cite{Hydrology}, oil recovery by water flooding~\cite{oilrecovery}, in enhancing the mixing of fluids~\cite{PhysRevLett.106.194502,Soltanian2017}, while fabricating structured soft materials~\cite{Marthelot2018} and in the control of dendritic growth morphologies in rechargeable batteries~\cite{dendrite}.  {The present work can provide useful insight on the transient behaviour of shear-thickening suspensions during the application of high shear rates, such as in impact behaviour~\cite{PhysRevE.97.052603}.} Our work also have useful implications in cementing processes involving the removal of drilling mud and its substitution with cement slurries~\cite{cement}.

\section*{CRediT authorship contribution statement}
\textbf{Palak:} Formal analysis, Software, Validation, Investigation, Writing
– original draft, Visualization, Data Curation. \textbf{Vaibhav Raj Singh Parmar:} Investigation, Formal analysis, Software, Data Curation, Writing – original draft, Validation, Visualization, Writing - Review \& Editing. \textbf{Sayantan Chanda:} Validation, Investigation, Writing - Review \& Editing. \textbf{Ranjini Bandyopadhyay:} Conceptualization, Methodology, Validation, Writing – Review \& Editing, Supervision, Funding acquisition, Project administration.

\section*{Declaration of Competing Interest}
The authors declare that they have no known competing financial interests or personal relationships that could have appeared to influence the work reported in this paper.

\section*{Data Availability}
Source data are available for this paper from the corresponding author upon reasonable request.

\section*{Acknowledgments}	 
 We thank Raman Research Institute (RRI, India) for funding our research and Department of Science and Technology Science and Education Research Board (DST SERB, India) grant EMR/2016/006757 for partial support.

\bibliographystyle{MS_style}
\bibliography{MS_ref}

\end{document}


	
	
\renewcommand{\figurename}{Supplementary Fig.}
	\setcounter{table}{0}
	\renewcommand{\thetable}{S\arabic{table}}%
	\renewcommand{\tablename}{Supplementary Table}
	\setcounter{figure}{0}
	\makeatletter 
	\renewcommand{\figurename}{Supplementary Fig.}
	\setcounter{figure}{0}
	\makeatletter 
	\renewcommand{\thefigure}{S\arabic{figure}}
	\setcounter{section}{0}
	\renewcommand{\thesection}{ST\arabic{section}}
	\setcounter{equation}{0}
	\renewcommand{\theequation}{S\arabic{equation}}
	\title{\color{blue}\textbf{\underline{Supplementary Information}\\ Emergence of transient reverse fingers during radial displacement of a shear-thickening fluid}}

	\author[1, $\dagger$, $\P$]{Palak}
	\affil[1]{\textit{Soft Condensed Matter Group, Raman Research Institute, C. V. Raman Avenue, Sadashivanagar, Bangalore 560 080, INDIA}}
	\author[1, $\ddagger$, $\P$]{Vaibhav Raj Singh Parmar}
        \author[1, $\S$]{Sayantan Chanda}
	\author[1,*]{Ranjini Bandyopadhyay}
	\date{\today}
 
	\footnotetext[5] {Palak and Vaibhav Raj Singh Parmar contributed equally to this work.}
	\footnotetext[2]{palak@rri.res.in}
	\footnotetext[3]{vaibhav@rri.res.in}
       \footnotetext[4]{sayantanc@rrimail.rri.res.in}
	\footnotetext[1]{Corresponding Author: Ranjini Bandyopadhyay; Email: ranjini@rri.res.in}
	\maketitle
	\pagebreak
		\definecolor{red(ryb)}{rgb}{1.0, 0.15, 0.07}
		\newcommand{\blsquare}{\textcolor{black}{\small$\blacksquare$}}
			\newcommand{\rcircle}{\textcolor{red(ryb)}{\large$\bullet$}}
		
  \begin{figure}[H]
    \centering
    \includegraphics[width=1\textwidth]{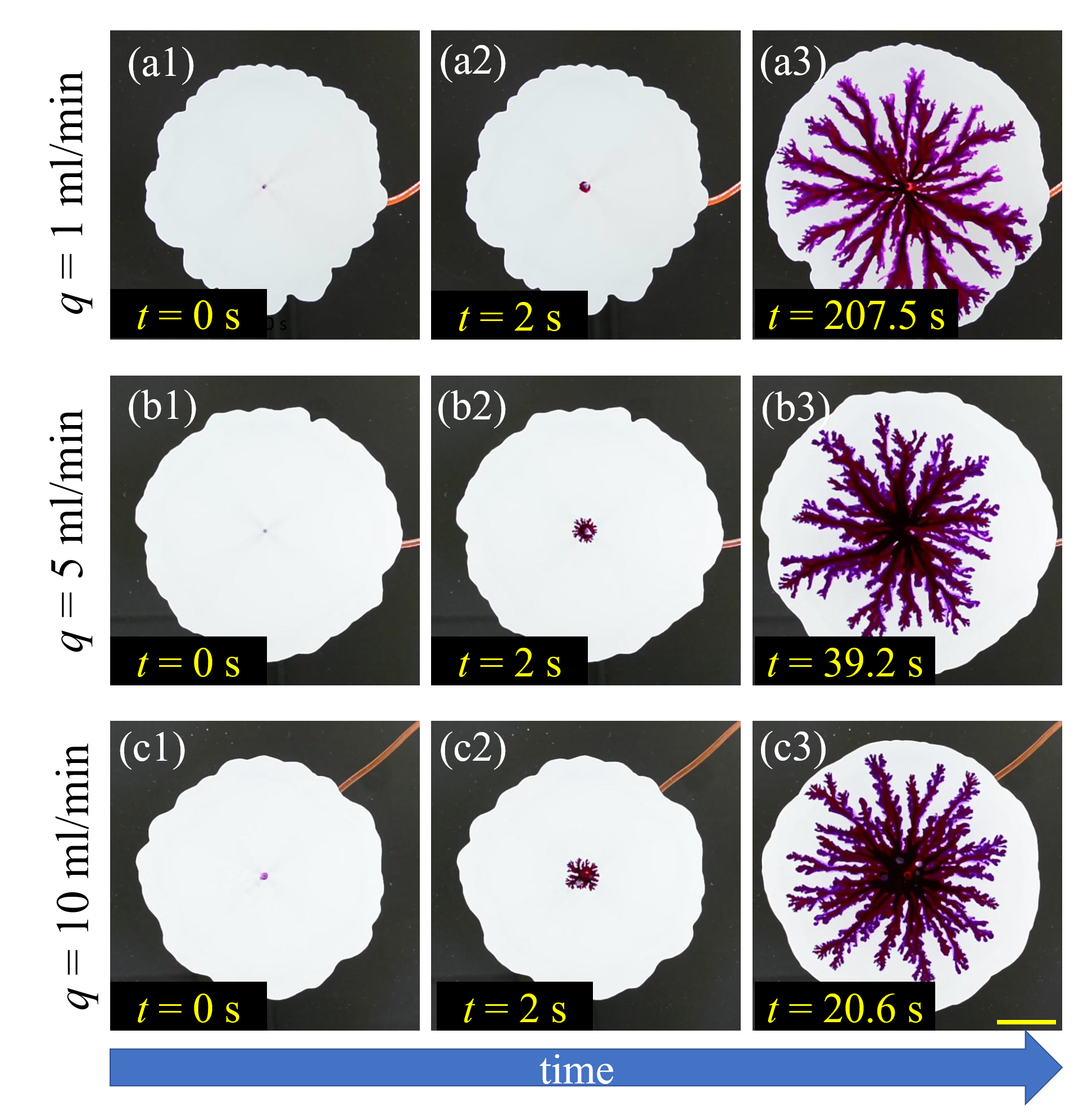}
    \caption{ Raw RGB images showing the temporal evolution of interfacial patterns during radial displacements of an aqueous 40 wt.\% cornstarch suspension by water at different flow rates: (a1-a3) $q$ = 1 ml/min (b1-b3) 5 ml/min (c1-c3) 10 ml/min.
The scale bar is 5 cm. The HS cell gap $b$ in these experiments is kept fixed at 170 $\mu m$. Transient withdrawal of the cornstarch suspension and the formation of reverse fingers are absent for these injection flow rates.}
    \label{fig:R2Q1-1}
\end{figure}

\begin{figure}[H]
    \centering
    \includegraphics[width=1\textwidth]{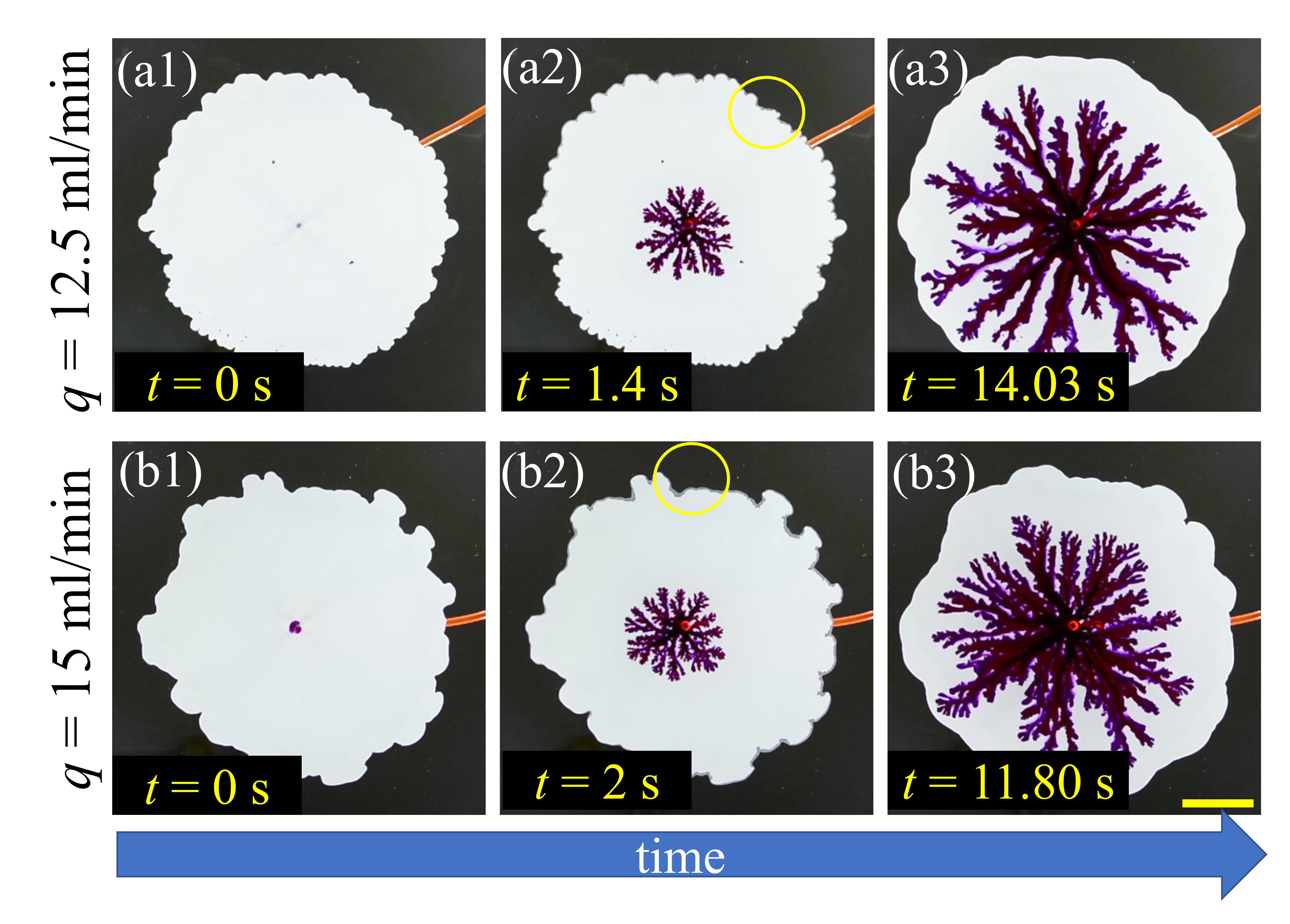}
    \caption{ Raw RGB images showing the temporal evolution of interfacial patterns during radial displacements of an aqueous 40 wt.\% cornstarch suspension by water at different flow rates: (a1-a3) $q$ = 12.5 ml/min (b1-b3) 15 ml/min. Yellow circles in (a2, b2) highlight the formation of transient reverse fingers at the time when they reach their maximum lengths.
The scale bar is 5 cm. The HS cell gap $b$ in these experiments is kept fixed at 170 $\mu m$.}
    \label{fig:R2Q1-2}
\end{figure}

\begin{figure}[H]
    \centering
    \includegraphics[width=0.85\textwidth]{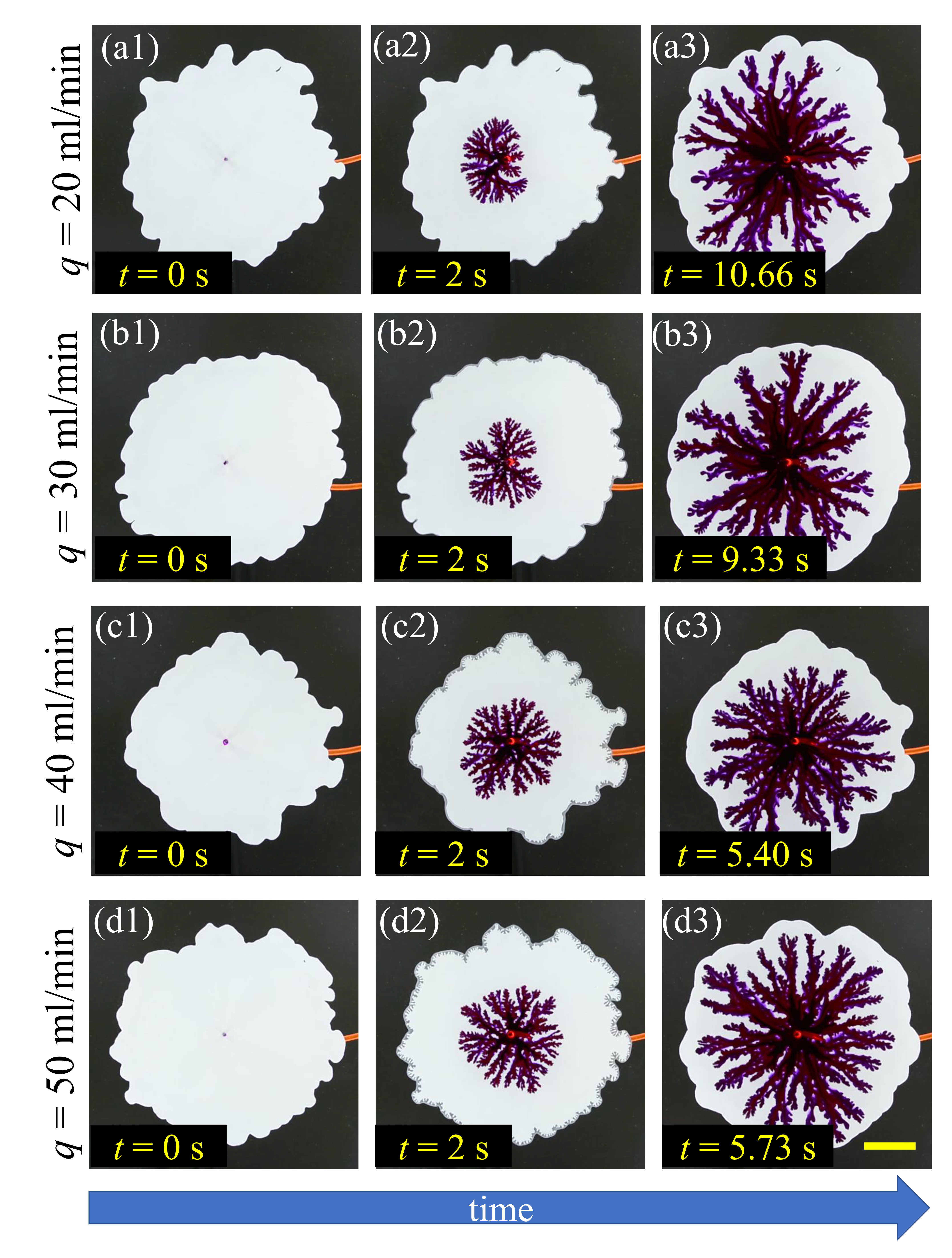}
    \caption{ Raw RGB images showing the temporal evolution of interfacial patterns during radial displacements of an aqueous 40 wt.\% cornstarch suspension by water at different flow rates: (a1-a3) $q$ = 20 ml/min, (b1-b3) 30 ml/min, (c1-c3) 40 ml/min, and (d1-d3) 50 ml/min. The scale bar is 5 cm. The HS cell gap $b$ in these experiments is kept fixed at 170 $\mu m$. Reverse fingers are observed in a2, b2, c2, d2, with their lengths increasing with increase in displacing flow rate.}
    \label{fig:R2Q1-3}
\end{figure}

    \begin{figure}[H]
	 	\includegraphics[width=1\textwidth]{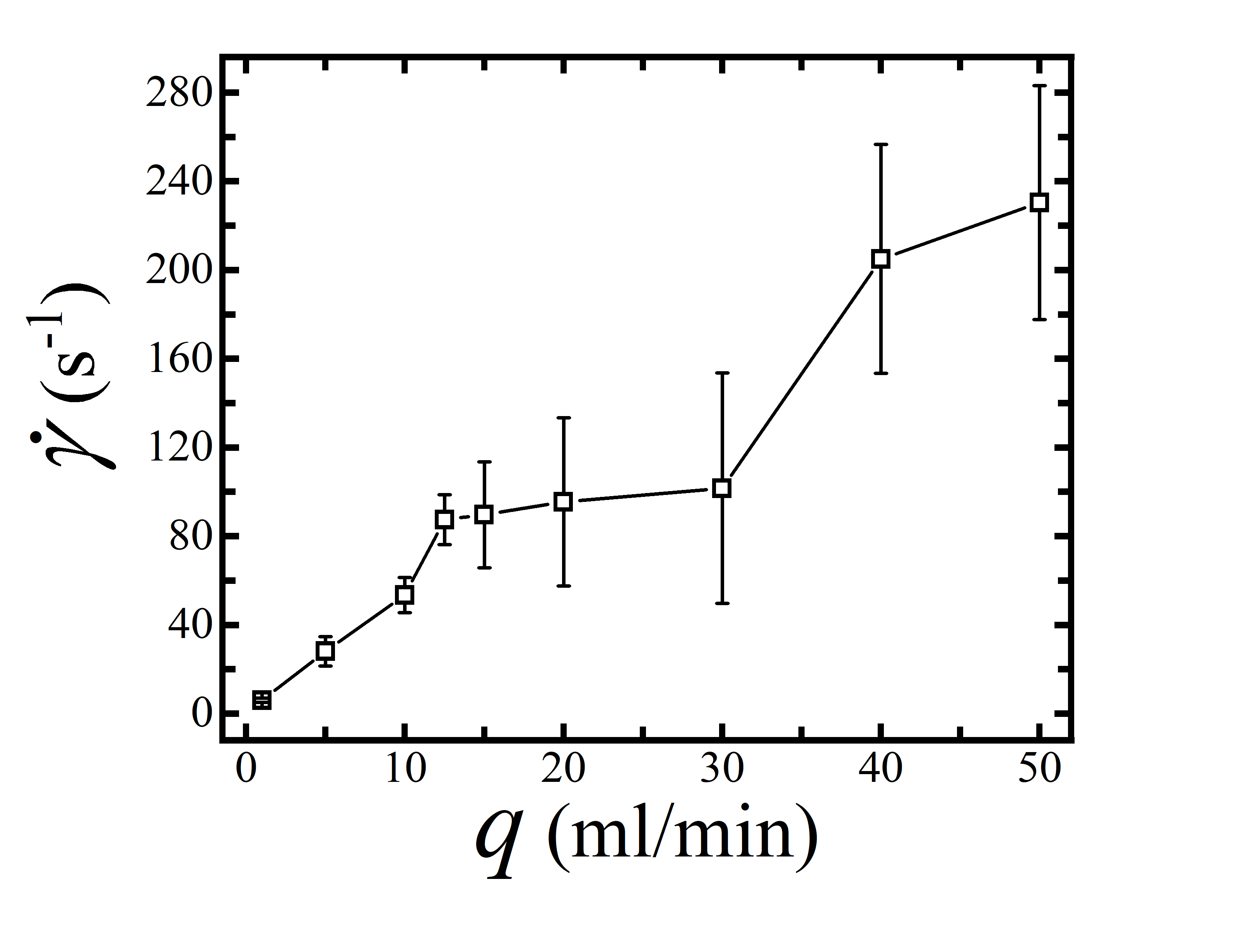}
	 	\centering
	 	\caption{\label{shearrate} Shear rates $\dot\gamma$ computed for various injection flow rates $q$ of the inner displacing fluid (water) during radial displacement of an aqueous 40 wt.\% cornstarch suspension in a Hele-Shaw cell of gap $b$ = 170 $\mathrm{\mu}$m. Error bars represent the standard deviations in measurements of shear rates for multiple tips.}
	 \end{figure}

 \begin{figure}[H]
	 	\includegraphics[width= 6.0in ]{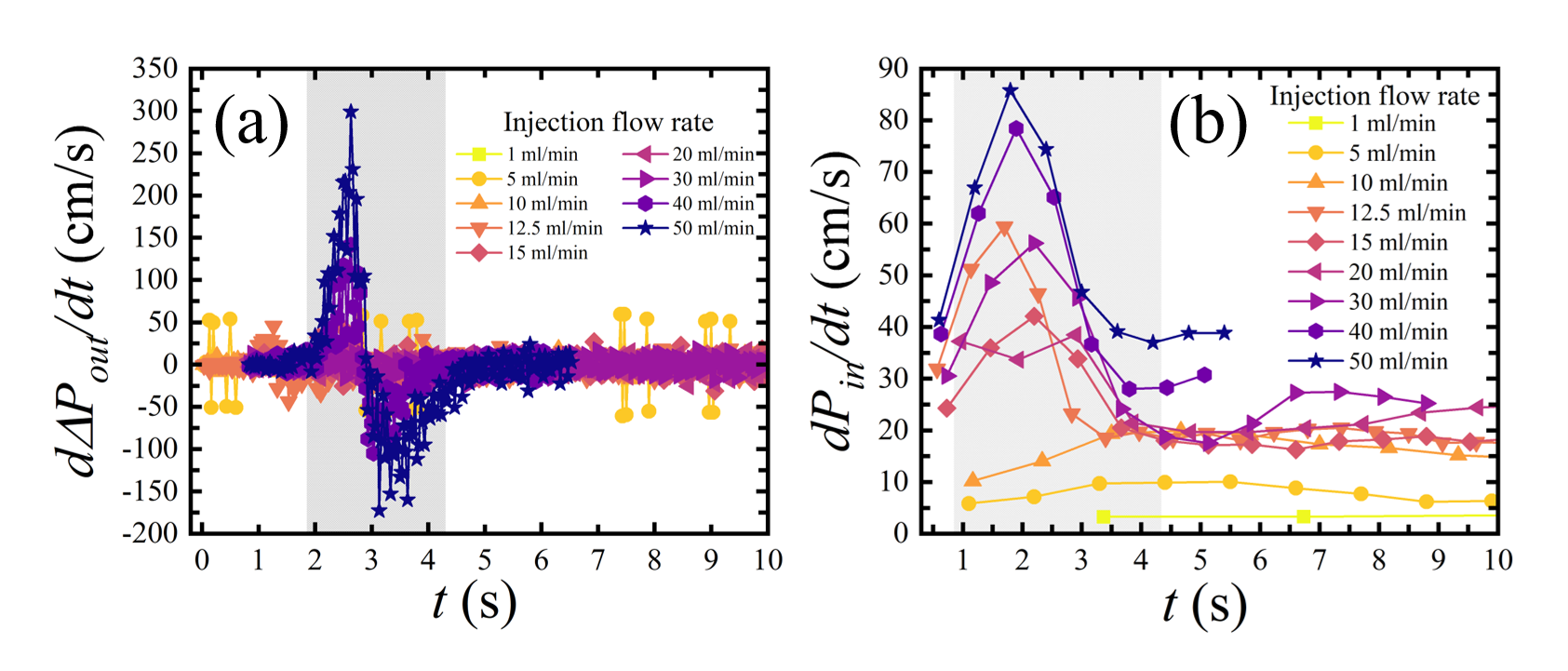}
	 	\centering
	 	\caption{\label{dPout}(a) $d\Delta P_{out}/dt$ vs. $t$ for various injection flow rates $q$. (b) $dP_{in}/dt$ vs. $t$ for various injection flow rates. The corresponding plots of $\Delta P_{out}$ and $P_{in}$ are shown in Fig.~3(a) of the main paper. The changes in the slopes of $d\Delta P_{out}/dt$ and $dP_{in}/dt$ occur during the same time interval (shown by grey shaded regions) and represent the underlying correlation in the growth kinetics of the inner and outer interfacial patterns. The HS cell gap in these experiments is kept fixed at $b$ = 170 $\mu$m and the concentration of the CS suspension is 40 wt.\%.}
	 \end{figure}
  
  \begin{figure}[H]
	 	\includegraphics[width= 5.0in ]{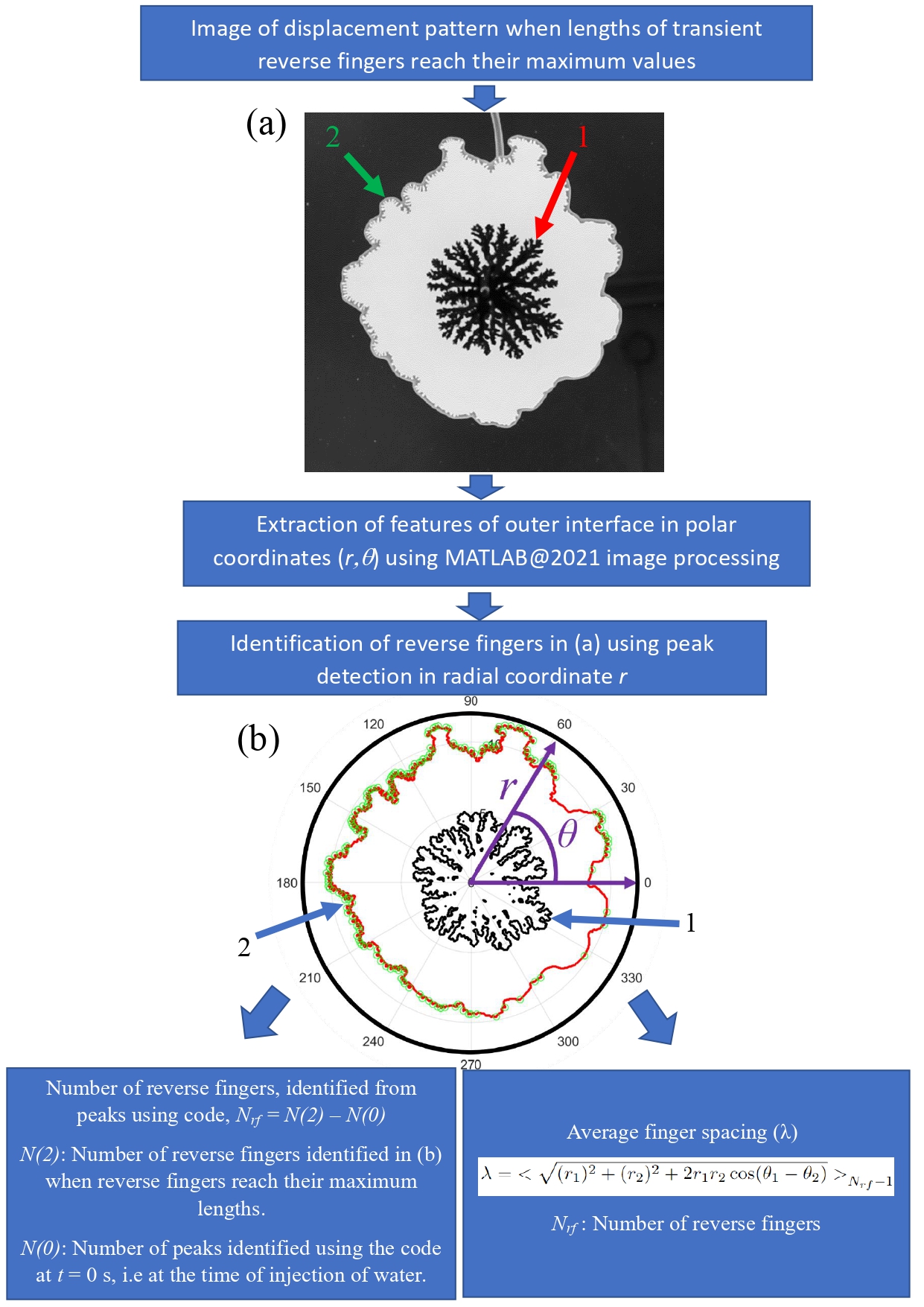}
	 	\centering
	 	\caption{\label{flowchart} Flow chart illustrating the procedure for calculation of the number of reverse fingers $N_{rf}$ and the average spacing $\lambda$ between them. Label 1 in (a-b) represents the inner interface between water and cornstarch suspension. Label 2 in (a-b) represents the outer interface between cornstarch suspension and air.}
	 \end{figure}

\begin{figure}[H]
    \centering
    \includegraphics[width=1\textwidth]{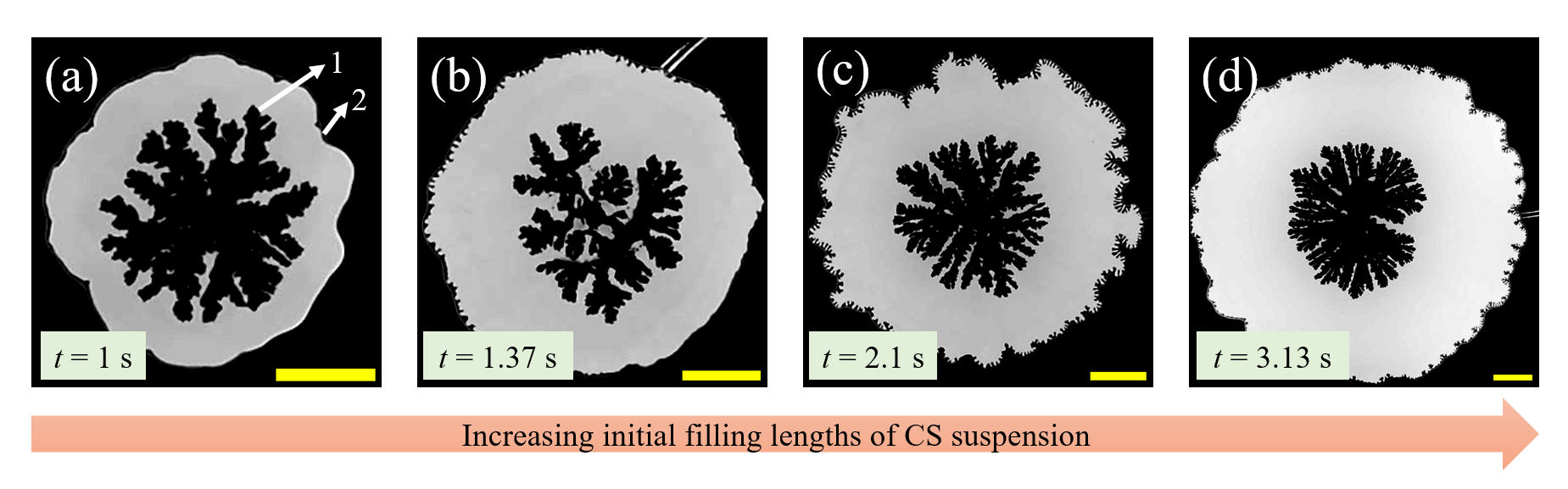}
    \caption{\textbf{Interfacial patterns for varying initial filling lengths of the outer cornstarch suspension} (a-d) Patterns in grayscale are shown at the time when transient withdrawal of the cornstarch suspension and the resultant reverse fingers reach their maxima. Initial filling lengths of the cornstarch suspension are (a-d) 5 cm, 7.5 cm, 10 cm and 15 cm respectively. The scale bar is 5 cm.}
    \label{fig:Picture3}
\end{figure}

\begin{figure}[H]
    \centering
    \includegraphics[width=1\textwidth]{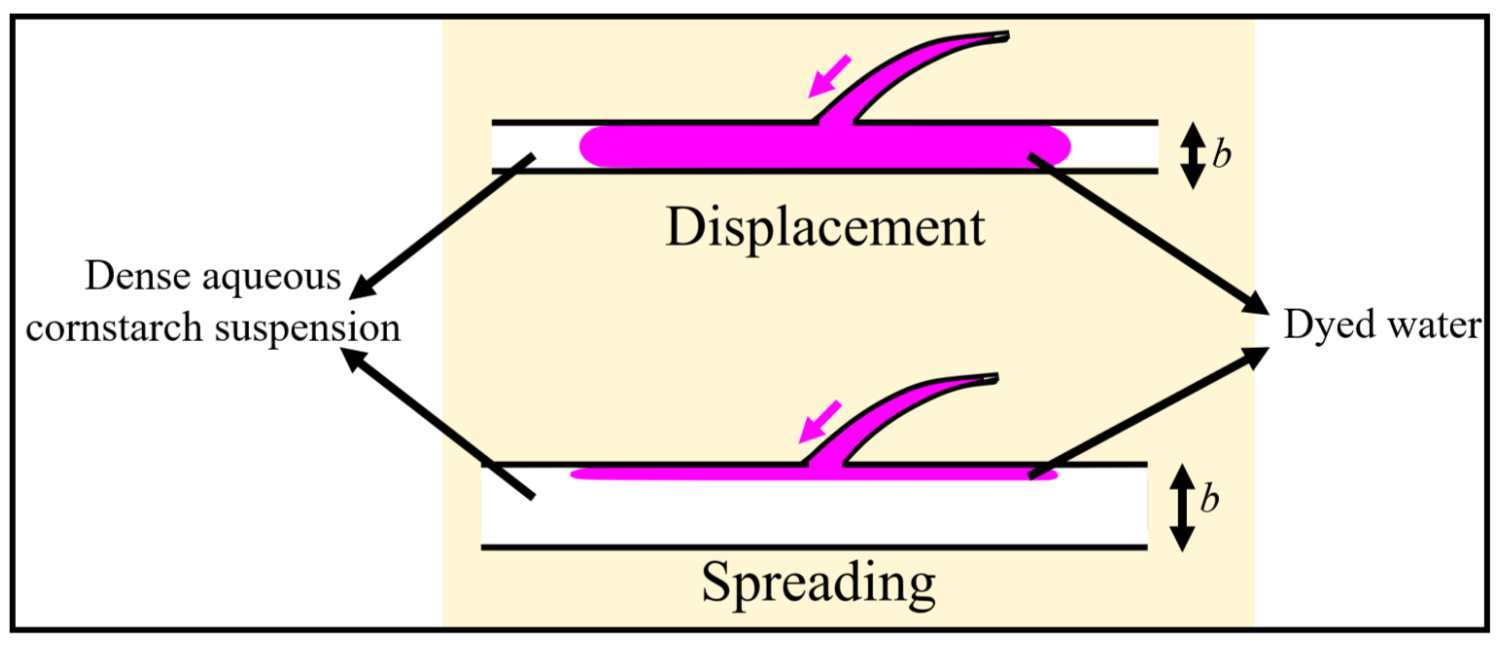}
    \caption{Schematic illustration of displacement vs. spreading. When water (pink) is injected into the Hele-Shaw (HS) cell, it displaces the cornstarch suspension (white) effectively across the HS cell gap only for very small gaps. In contrast, the spreading of water on the CS suspension is identified when displacement is not effective across the gap of the HS cell, which results in spreading rather than displacement for larger cell gaps.}
    \label{fig:spread}
\end{figure}

  \begin{figure}[H]
	 	\includegraphics[width=4.5in ]{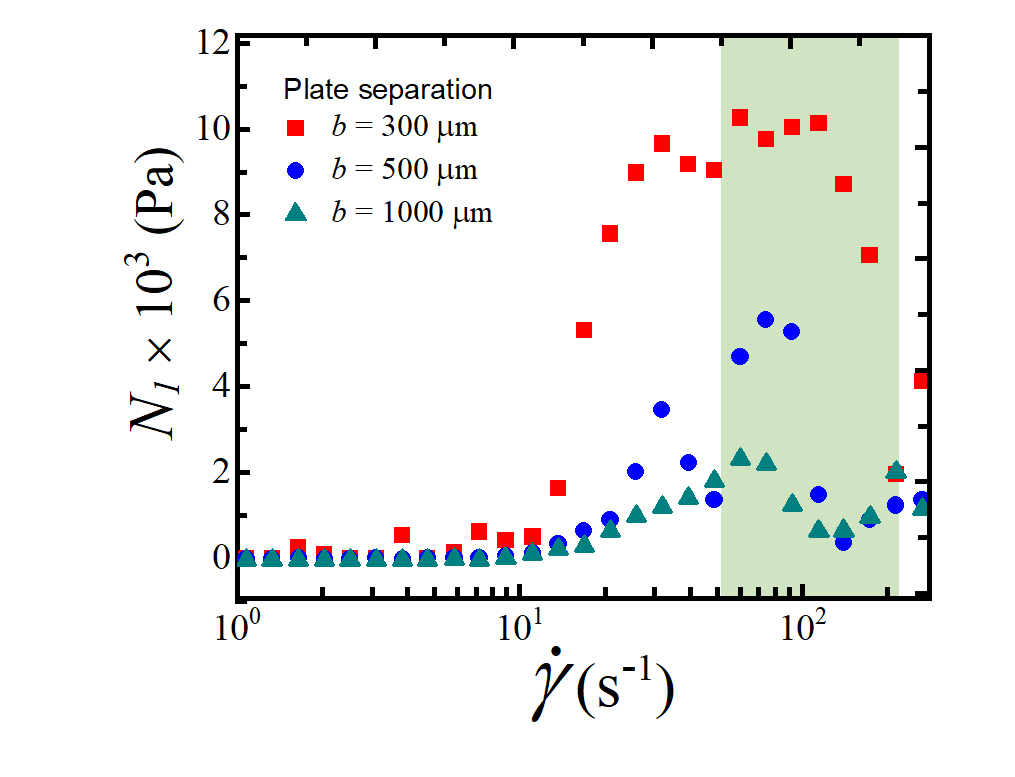}
	 	\centering
	 	\caption{\label{Normalgap} First normal stress difference $N_1$ vs. shear rate $\dot{\gamma}$ for a 40 wt.\% cornstarch suspension measured for different plate separations in the rheometer. The highlighted region indicates the shear rate range explored in the displacement experiments for various gaps of the HS cell.}
	\end{figure}

\begin{figure}[H]
    \centering
    \includegraphics[width=1\textwidth]{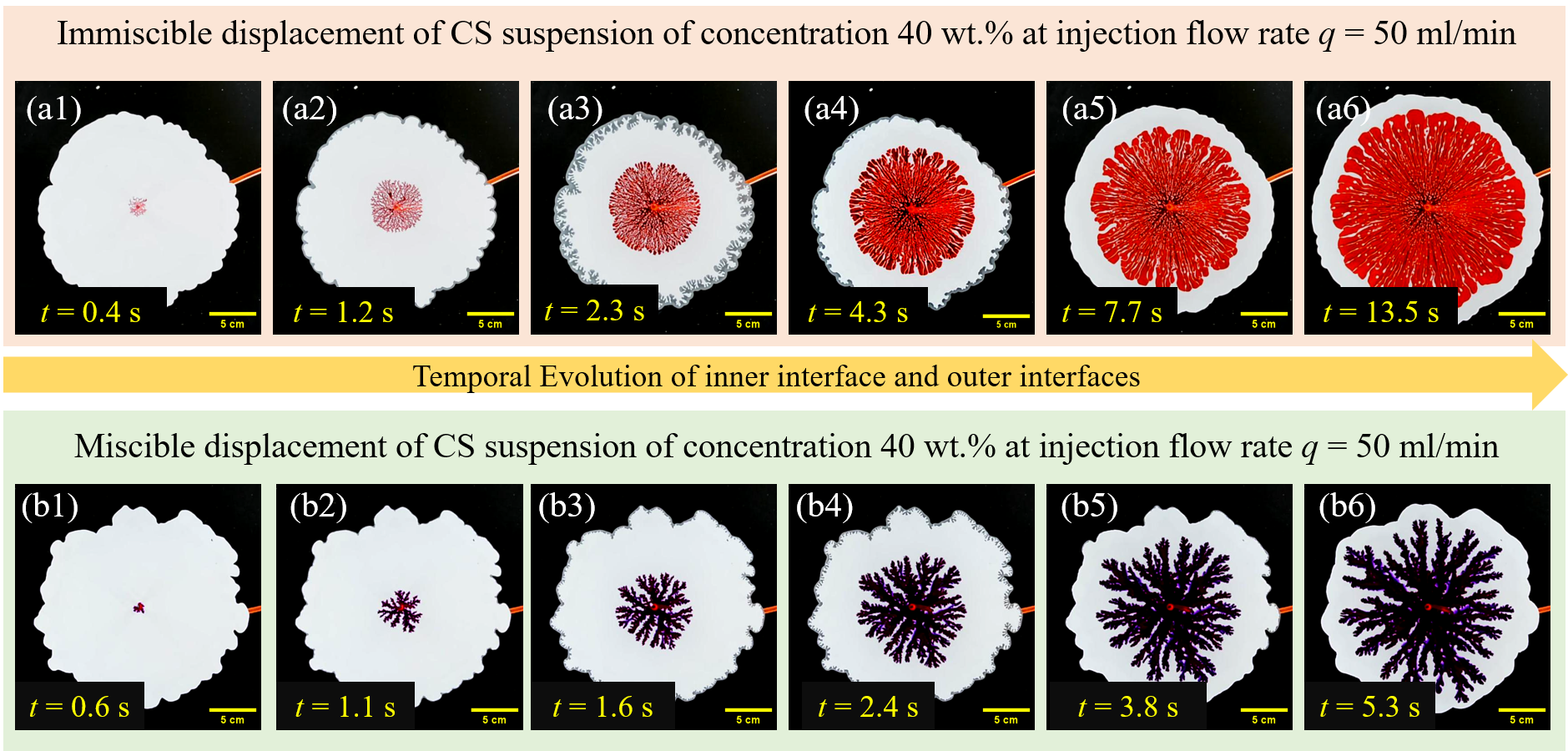}
    \caption{\textbf{Effect of miscibility of the fluid pair on the formation of reverse fingers at the outer interface:} Temporal evolution of inner and outer interfaces when a CS suspension of concentration 40 wt.\% is displaced by (a1-a6) immiscible mineral oil and (b1-b6) miscible water at an injection flow rate $q$ = 50 ml/min. The scale bar is 5 cm. The reverse fingers reach their maximum lengths at $t$ = 2.4 s and 2.1 s for immiscible and miscible displacement experiments respectively.}
    \label{fig:immiscible}
\end{figure}

 \begin{figure}[H]
	 	\includegraphics[width= 6.5in ]{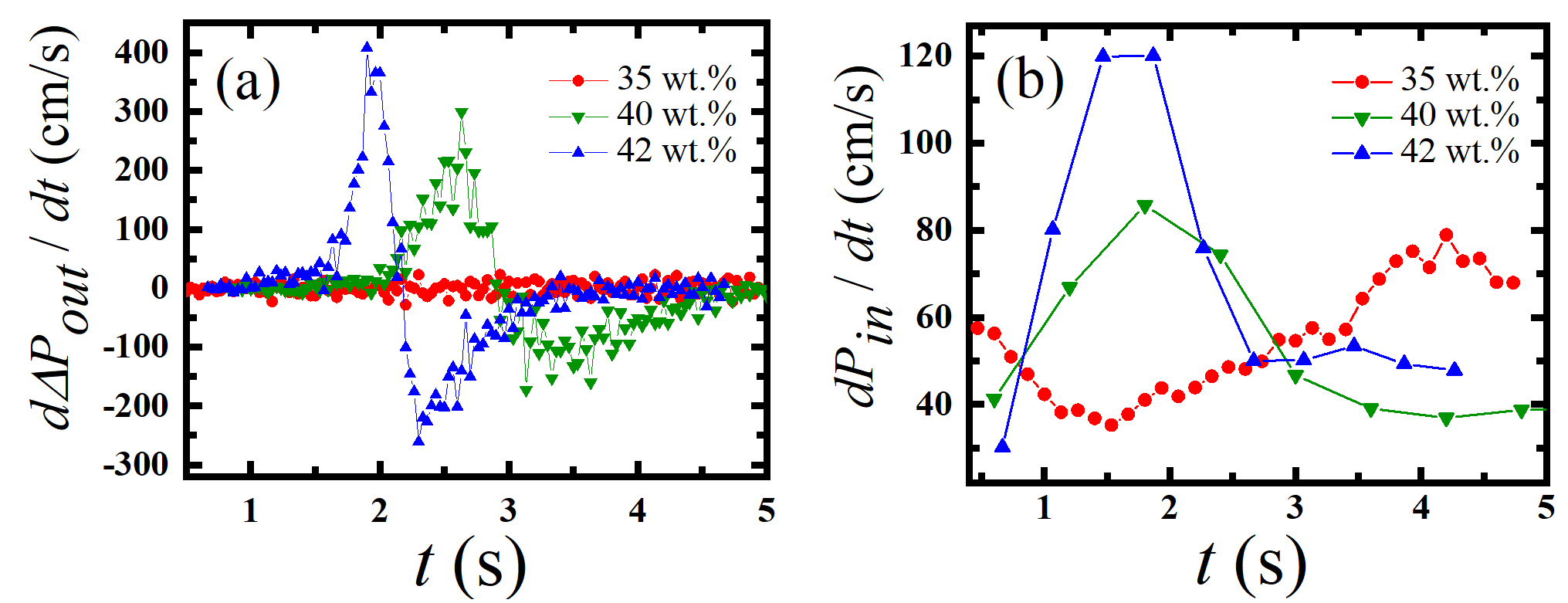}
	 	\centering
	 	\caption{\label{Varconc} (a) $d\Delta P_{out}/dt$ vs. $t$ and (b) $dP_{in}/dt$ vs. $t$ obtained during the displacements of cornstarch suspensions of different concentrations by water at a fixed injection flow rate $q$ = 50 ml/min for a HS cell gap $b$ = 170 $\mu$m.}
\end{figure}